%% file: main.tex
\documentclass{article}

\usepackage{arxiv}

\usepackage[utf8]{inputenc} % allow utf-8 input
\usepackage[T1]{fontenc}    % use 8-bit T1 fonts
\usepackage{hyperref}       % hyperlinks
\hypersetup{hidelinks}
\usepackage[hang,flushmargin]{footmisc} % flush-left footnotes
\usepackage{url}            % simple URL typesetting
\usepackage{booktabs}       % professional-quality tables
\usepackage{amsfonts}       % blackboard math symbols
\usepackage{nicefrac}       % compact symbols for 1/2, etc.
\usepackage{microtype}      % microtypography
\usepackage{lipsum}
\usepackage{fancyhdr}       % header
\usepackage{graphicx}       % graphics

\usepackage{macros}
\usepackage[numbers,sort&compress]{natbib}
\usepackage{xurl}
\usepackage{hyperref}
\hypersetup{hidelinks}

\usepackage{tabularx}
\usepackage{lscape} 
\usepackage{enumerate}
\usepackage{amsmath}
\usepackage[inline]{enumitem}

\usepackage{pgfplots} 
\pgfplotsset{compat=1.18}

\AtBeginDocument{
  \fancypagestyle{firstpagestyle}{
    \fancyhead{}%
    \fancyfoot[C]{}%
  }
  \fancyhf{}
\pagestyle{fancy}
\fancyhf{}

\newcommand{\runningtitle}{Methods and Motivations of AI-Generated Sexual Content Creators}
\newcommand{\runningauthors}{Mink, Qin, Redmiles}

\fancyhead[L]{\sffamily\footnotesize\ifodd\value{page}\runningtitle\fi}
\fancyhead[R]{\sffamily\footnotesize\ifodd\value{page}\else\runningauthors\fi}
\fancyfoot[C]{}

  \fancyfoot[C]{}%
}
\makeatother

\title{``Unlimited Realm of Exploration and Experimentation'': Methods and Motivations of AI-Generated Sexual Content Creators}

\author{
Jaron Mink\footnotemark[1]\\
Arizona State University \\
\texttt{jaron.mink@asu.edu}
   \And
  Lucy Qin\footnotemark[1] \\
  Georgetown University\\
  \texttt{lucy.qin@georgetown.edu} \\
     \AND
  Elissa M. Redmiles \\
  Georgetown University\\
  \texttt{elissa.redmiles@georgetown.edu} \\
}

\begin{document}
\maketitle

\begingroup
\renewcommand{\thefootnote}{}
\footnotetext[1]{* \textbf{Both authors contributed equally} to this work, share first authorship, and reserve the right to adjust ordering on personal documents.}

\footnotetext{This work appears in the proceedings of {the 2026 ACM Conference on Fairness, Accountability, and Transparency (FAccT)}. This extended version includes additional analyses and discussion.}
\endgroup

\begin{abstract}
\input{abstracts/abstract}

\end{abstract}

\vspace{1.5\baselineskip}

\noindent\fbox{%
  \parbox{0.98\columnwidth}{%
    {\textbf{Content Warning:} 
    This paper discusses and contains quotes describing sexual content. This work also discusses sexual violence, CSAM, violence, gore, and identity-based harassment, and includes first-hand accounts from people who have perpetrated image-based sexual abuse by creating AI-generated intimate imagery that nonconsensually depicts an identifiable individual. 
    }
  }
}

\newpage

\input{arXiv-sections/intro}
\input{arXiv-sections/related}

\input{arXiv-sections/methods}
\input{arXiv-sections/background-and-content}
\input{arXiv-sections/use-cases}

\input{arXiv-sections/discussion}

\section*{Acknowledgments}
This work was supported in part by NSF Award 2513313. We thank Rosanna Bellini, David A. Forsyth, \new{Vaughn Hamilton}, Tadayoshi Kohno, Mantas Mazeika, \new{Allison McDonald,} Aurélie Petit, Filipo Sharevski, Zahra Stardust, Gianluca Stringhini, Sharon Wang, Miranda Wei, and Eric Zeng for assistance distributing recruitment materials, feedback on earlier drafts, and/or providing advice on researcher protection. We also thank community moderators for their feedback.

%Bibliography
%\bibliographystyle{unsrt}

\bibliographystyle{ACM-Reference-Format}
\bibliography{references/bib-strings, references/polished-refs}

\newpage
\appendix
\input{appendix}

\end{document}

%% file: abstracts/abstract.tex
AI-generated media is radically changing the way content is both consumed and produced on the internet, and in no place is this potentially more visible than in sexual content. AI-generated sexual content (AIG-SC) is increasingly enabled by an ecosystem of individual AI developers, specialized third-party applications, and foundation model providers. AIG-SC raises a number of concerns from older debates about the line between pornography and obscenity to newer debates about fair use and labor displacement (in this case, of sex workers), and has spurred new regulations to curb the spread of non-consensual intimate imagery (NCII) created using the same technology used to create AIG-SC. However, despite the growing prevalence of AIG-SC, little is known about its creators, their motivations, and what types of content they produce. To inform effective governance in this space, we conducted an in-depth study to understand what AIG-SC creators make, along with how and why they make it. Interviews with 28 AIG-SC creators, ranging from hobbyists to entrepreneurs to those who moderate communities of hundreds of thousands of other creators, revealed a wide spectrum of motivations, including sexual exploration, creative expression, technical experimentation, and in a handful of cases, the creation of NCII.

%% file: arXiv-sections/intro.tex
\newpage
\setcounter{footnote}{0}
\section{Introduction}
Sexual content\footnote{We define \textit{sexual content} as an information medium (image, video, text, or audio) that depicts sexual acts, erogenous zones, or is intended to be sexually arousing to the viewer, and includes erotic images, fiction, or roleplaying, among others.} has long played a central role in the adoption of technology; from home video to the internet \revision{and cryptocurrencies, sexual material financially supports and makes early use of emerging technology}~\cite{stardustHighRiskHustling2023}.
Generative AI now stands as the next frontier; 
\revisiontwo{with AI, users can now make custom sexual images, audio, video, and text~\cite{weatherbedXAIsNewGrok2025, edwardsChatGPTCanNow2025}.} 
This AI-generated sexual content (AIG-SC)\footnote{We use the following acronyms to refer to different types of AI-generated content. \textbf{AIG-NCII} refers to AI-generated \textbf{non-consensually created intimate imagery} (sometimes referred to as ``sexualized deepfakes'' or ``deepfake pornography'') that depicts a known individual's likeness. \textbf{AIG-CSAM} refers to AI-generated \textbf{child sexual abuse material}. \textbf{AIG-NCC} refers to AI-generated \textbf{non-consensual content}, which we use as an umbrella term to include both AIG-NCII and AIG-CSAM. Lastly, \textbf{AIG-SC} refers to AI-generated \textbf{sexual content} (sometimes referred to as ``AI pornography'') that does not include AIG-NCC.} ecosystem is becoming increasingly supported by third-party applications for local~\cite{coheeSillyTavern2025, comfyorgComfyUI2025} and cloud-based~\cite{pixaiPixAIWebsite} generation, and now even by mainstream foundation model providers like OpenAI \revision{and xAI} via age-verified accounts~\cite{france-presseOpenAIWillAllow2025, weatherbedXAIsNewGrok2025}. \new{Large communities devoted to AIG-SC exist and continue to grow. For example, an active Reddit community that is associated with a broader AIG-SC group~\cite{kylewiggersMeetUnstableDiffusion2022} has over 300,000 members.}

\new{In response, researchers are increasingly trying to understand what positive and negative effects AIG-SC has on society.
In addition to a number of mixed findings on the positive and negative impacts AIG-SC has upon \textit{consensual} users and stakeholders of AIG-SC systems~\cite{lafortune2026could,hightow2022tough,willoughby2025artificial}, 
the prevalence and deep harms of \textit{non-consensual} AIG-SC are only beginning to be understood.}
The technology powering AIG-SC significantly scales sexual abuse through the creation of AI-generated non-consensual content (AIG-NCC): specifically, non-consensual intimate imagery (AIG-NCII), which as defined by the newly passed U.S. TAKE IT DOWN Act~\cite{tedcruzS146TAKEIT2025} is an ``intimate visual depiction of an identifiable individual... that, when viewed as a whole by a reasonable person, is indistinguishable from an authentic visual depiction of the individual'', and child sexual abuse material (AIG-CSAM)~\cite{fbiChildSexualAbuse2024}, ``any visual depiction of sexually explicit conduct involving a person less than 18 years old.'' AIG-NCC is sexual abuse~\cite{internetwatchfoundationHowAIBeing2023,henryImagebasedSexualAbuse2021}. Such content is a harmful violation, and research has documented the multitude of severe downstream consequences experienced by victim-survivors, including suicide and PTSD~\cite{henryImagebasedSexualAbuse2021,thorn&atihThornSafetyDesign2024}. 

In the face of the growing cases of AIG-NCC, the attorneys general of 48 U.S. states have called upon platforms and payment processors to assist in reducing this content~\cite{generalAGSundayLeads2025}. Some of the ensuing actions were effective and targeted, such as revoking service from platforms hosting AIG-NCII models~\cite{maibergCreditCardCompanies2025} and shutting down AIG-NCII marketplaces~\cite{wiseMajorDeepfakePorn2025}, while others were broad: companies updating usage policies to restrict general generation of sexual content~\cite{googleGeminiAppPolicy2025, edwardsChatGPTCanNow2025, aiAcceptableUsePolicy2025}, delisting of AIG-SC in search results~\cite{albaGoogleWorksReduce2024}, and removing any sexual content from online marketplaces~\cite{maibergCreditCardCompanies2025}. These latter broad actions, affecting all AIG-SC rather than only AIG-NCC, have triggered concerns from digital rights groups about the restriction of individuals' expression and privacy rights~\cite{mullinTAKEITAct2025, centerfordemocracy&technologyTAKEITAct2025}.

Addressing complex situations such as this tension between free speech (AIG-SC) and abuse (AIG-NCC) requires effective governance~\cite{scottSeeingStateHow2020}, which in turn requires structured grounding in the existing situational context~\cite{ostromBackgroundInstitutionalAnalysis2011}: an understanding of the actors, their actions, and why they are taken. \revision{Prior work by security and privacy scholars~\cite{gibsonAnalyzingAINudification2025, hanCharacterizingMrDeepFakesSexual2025, brighamViolationMyBody2024, umbachNonconsensualSyntheticIntimate2024, wei2025utterlyillprepared} has understandably focused on studying AIG-NCC (see Section~\ref{subsect:related-work-NCII} for a detailed review). Beyond this work on AIG-NCC, little is known about AIG-SC. Protecting people from abuse while also protecting free speech will require an understanding of the ecosystem surrounding AIG-SC use cases as well.} To establish foundational context for discussions on the governance of AIG-SC, we ask the following:

\begin{enumerate}[leftmargin=*,label=\textbf{RQ\arabic*}]
\item \textit{Who} creates AIG-SC and what initiates their creation?\label{rq:background}
\item \textit{What} types of AIG-SC do creators make and \textit{how}?\label{rq:content}
\item \textit{Why} do creators generate AIG-SC?\label{rq:motivations}
\end{enumerate}

We answer these questions through semi-structured interviews with 28 AI-generated sexual content (AIG-SC) creators. Our participants ranged from hobbyists, to professional jailbreakers, to sex workers, to tool builders and community moderators whose resources are used by as many as hundreds of thousands of creators. 
While we explicitly recruited from communities with policies against AIG-NCC, our sample included three people who, in the course of their interview, disclosed making both AIG-NCC (specifically, AIG-NCII) and AIG-SC, reflecting the reality of existing ecosystems in which some AIG-SC communities include people who create both AIG-NCC and AIG-SC.
\revisiontwo{To conduct this research ethically and responsibly, the research team followed established best practices for interviewing perpetrators of sexual violence, and engaged in continuous discussion internally and with researchers from non-CS fields that have established norms and a history
of engagement in research on perpetrators of sexual violence; we detail our procedures and considerations in Section~\ref{sec:ethics}.}
\revision{As a result of these encounters, we added, and answer, the following research question:}
\begin{enumerate}[start=4,leftmargin=*,label=\textbf{RQ\arabic*},topsep=3pt]
\item \new{How do perpetrators who inhabit AIG-SC spaces create and rationalize AIG-NCC?} \label{rq:AIG-NCC}
\end{enumerate}

\header{\new{Positionality Statement}}
Our team consists of researchers who have studied sexual content in the context of sex work, image-based sexual abuse, and synthetic content creation and moderation for more than 5 years. The research team holds the following relevant beliefs and experiences: the team takes a sex-positive stance in our research on sexual content and interactions; the team supports the decriminalization of sex work; the team includes members who have experienced sexual abuse; the team strongly believes in both the prevention of sexual violence and the protection of free speech, including sexual expression. None of the team actively participates in an AIG-SC community.
We conduct this work to enable legal, platform, and community-based governance that actively prevents sexual violence through further understanding of AIG-SC and its delineations from AIG-NCC.

\header{Summary of Findings} 

\noindent(\ref{rq:background}) Many participants came from computing backgrounds, but also from the arts and the sex industry. They may have discovered AIG-SC through encountering a safety filter on a general-purpose generative AI (GenAI) tool that piqued their curiosity, \revision{a niche sexual or creative interest, innate curiosity, or in pursuit of a highly engaged community to teach them general AI skills.}

\smallskip
\noindent 
(\ref{rq:content}) Participants primarily created text-based and image-based content; some created audio and several mentioned interest in video generation once capabilities improve. Participants created content using: (1) customized pipelines---custom compositions of prompting tools, generative models, fine-tuning layers such as low-rank adaptations (LoRAs), and sometimes manual modification (e.g., using Adobe Photoshop) strung together by a scripting language like Python; (2) UI-bound approaches---direct UI-based prompting of task-specific services like character.ai or proprietary general-purpose generative models, including mispurposing using jailbreaks; and/or (3) commissions---hiring someone to create or improve their content. Similar to studied ecosystems of cybercrime and abuse~\cite{tsengToolsTacticsUsed2020}, technically savvy community members upskilled others, creating an inviting space for learning AI skills specialized to sexual content. 

 \smallskip
\noindent 
(\ref{rq:motivations})  Participants described making content to satisfy niche sexual interests, express themselves sexually or creatively, challenge themselves technically and/or learn new skills, make money, gain membership in a community where they were free to discuss their sexual interests or get technical mentorship, and/or challenge perceived sexual censorship. In Section~\ref{sec:discussion} we discuss the need for further research on the impact of AIG-SC on society.

\smallskip
\noindent 
(\ref{rq:AIG-NCC}) 
Three participants (out of 28) disclosed that they intentionally created AIG-NCC (specifically, NCII) and one \revision{other} participant mentioned that a friend did so. To create AIG-NCC of people they knew, celebrities, and strangers, they used, individually and in combination, nudification services, face-swapping algorithms, and overlapping with AIG-SC techniques, jailbreaking and fine-tuning of machine learning models. They were motivated by sexual gratification, curiosity, community status, and compensation. Some justified their behavior as causing no harm, consistent with prior work, while others justified their behavior under the auspices of security.
In Section~\ref{sec:discussion} we discuss the need for, and potential directions toward achieving, community-based intervention against this abusive and harmful behavior taking place in AIG-SC communities that have stated policies against it. In so doing, we build on prior work on hacker communities and the factors that drive progression toward white- vs. black-hat activities.
We conclude with a call for computer scientists to consider their role in this space given the prevalence of technical employment and training among AIG-SC creators.

%% file: arXiv-sections/related.tex
\section{Background and Related Work}
\label{sec:relatedwork}
\subsection{Historical Perspectives}
\revisiontwo{Spurred by the emergence of new media and technology, questions over how to both legally define sexual content and practically regulate abusive material are not unique to AI. We briefly provide some historical and terminological context on when these same questions were raised over the past half-century \facct{within the U.S.}}

\header{The Porn Wars} 
The current legality of digital sexual content in the United States, and its protection under free speech laws, is the result of a large-scale legal and social battle held in the 1980s-1990s known as the ``Porn Wars''~\cite{fullerPornWarsSerious2019}.
Incited from a schism around sexuality within the feminist movement~\cite{wilson1983context}, a coalition of anti-porn feminists and conservative Christians stood at odds against pro-porn and anti-censorship feminists,
arguing the value of porn as sexual expression.

As lawyers battled over whether pornographic content was obscene under the precedent set by the Miller Test,\footnote{Based on precedent set by United States Supreme Court ruling on Miller v. California (1973), speech is considered obscene and not legally protected if an average person finds the material to be (1) of prurient interest, (2) offensive, and (3) lacking literary, artistic, political, or scientific value~\cite{u.s.supremecourtMarvinMillerState1973}.} social arguments focused on:
\textit{(1) Impact on Viewers (and Society):} 
whether viewers were impacted in ways that are personally or societally harmful.
Anti-porn activists argued that pornography was both a systematic cause and result of gender-based discrimination~\cite{mackinnon1985pornography}; intersecting with portrayals of battery, rape, sexual harassment, prostitution, and child abuse, it was argued that sexual violence was implicitly encouraged~\cite{mackinnon1985pornography, cowan1988dominance, palys1986testing, yang1990movie}. 
Pro-porn and anti-censorship activists argued that pornography (and particularly content that is non-normative, kinky, and queer) enabled the challenge of gender roles and stereotypes~\cite{strossen1993feminist} and that censorship of pornography would undermine the rights of women and queer individuals~\cite{strossen1993feminist}. Empirically, the impact of pornographic content on attitudes, beliefs and behavior is unclear: some systematic reviews find that pornography use correlates with non-sexual violence~\cite{mestre-bachPornographyUseViolence2024}, others find that sexual offenders had \textit{less} exposure to pornography than non-offenders~\cite{mellorUsePornographyRelationship2019}. Studies have failed to establish a causal relationship between pornography consumption and sexually violent beliefs and behaviors~\cite{mestre-bachPornographyUseViolence2024,grubbsPornographyUsePsychological2021}, while positive effects of pornography use have been demonstrated, although such effects are understudied~\cite{grubbsPornographyUsePsychological2021, doring2009internet}. In sum, a 2023 review concludes, ``there is limited evidence that pornography use always or even consistently leads to inherently negative outcomes, but, rather, its effects seem variable depending on a range of individual and sociocultural factors''~\cite{grubbsPornographyUsePublic2023}.

\textit{(2) Safety of Women in Pornography:}
While all sides are firmly against the production of NCII, definitions of consent and coercion significantly differ.
In the shadow of high-profile reports of abuse and coercion in pornography~\cite{lovelace2005ordeal},
anti-pornography activists argued that sexual depictions of women were inherently exploitative and coercive, making consent impossible to give~\cite{baldwin1984sexuality}. Pro-pornography activists not only critique these stances as paternalistic and broadly undermining of women's rights,
but argue that censorship would only drive black-market production, further endangering women who could otherwise be legally protected~\cite{strossen1993feminist}.

While courts ultimately recognized and protected sexual privacy and freedom of fantasy,
sexual material can still be deemed as obscene, often in line with prejudices against queer and non-heteronormative content (see ``\textbf{Sexualities \& Sex Positivity}''). Before the 1980s, pornographic content was often shown and viewed in adult cinemas, allowing for censor boards to pre-review and approve films~\cite{kleinhansChangeFilmVideo2006}. Major technological advances in the following decades -- at-home video, the early internet, creator economies, and now, AI -- have consistently reinvigorated debates about the regulation of the technology used to create and share sexual content, particularly as these advances decrease visibility into what individuals do in the privacy of their computing environments. 

\header{Sexualities \& Sex Positivity}
Sexuality~\new{\cite{APASexuality}} refers to all aspects pertaining to one's sexual behavior, including identities relating to one's gender (e.g., non-binary: an umbrella term for genders that do not fall within the strict social categories of a man or woman), sexual orientation (e.g., pansexual: an attraction to all genders), or activities (e.g., kinky: engaging in unconventional sexual activities).
To this day, heteronormative beliefs (the assumption of heterosexuality as normal or natural~\cite{americanpsychologicalassociationHeteronormativity}) dominate society, affecting policies, science, medical care, and, importantly, legal interpretations of obscenity. 
For instance, obscenity law has been used by the CDC to bar AIDS-educational content that ``promote, or encourage, homosexual activities''~\cite{u.s.districtcourtforthesoutherndistrictofnewyorkGayMensHealth1989}, 
and prevent the distribution of lawful abortion-inducing medication~\cite{beckAntiAbortionExtremistsWant2024}.
In response to stigma and censorship of safe consensual sexual activity,
a social movement known as ``sex-positivity'' has advocated for a recognition of sexuality as a human experience deserving of respect, acceptance of diverse sexualities, and ensuring access to sexual information and healthcare~\cite{ivanski2017exploring}. ``Sex-positivity'' \textit{is not} the encouragement of more sexual activity, or unconditional condoning of sexual acts~\cite{sexualityeducationresourcecentreSexPositivity2026}. Finding that they result in better patient outcomes, several professional fields, such as social work~\cite{williamsResolvingSocialProblems2013}, clinical and counseling psychology~\cite{williamsResolvingSocialProblems2013, sprott2023clinical, herbitter2024mental, pillai-friedmanBecomingKinkawareaNecessity2015, alexander2019sex}, and general healthcare~\cite{nimbi2022biopsychosocial}, have begun incorporating sex-positive approaches. 

\revision{Furthermore, sex-positive approaches have begun to be used in reducing sexual trauma for victim-survivors and perpetration of sexually violent attitudes and behaviors.}
Multiple studies of NCII victims~\cite{walkerSystematicReviewCurrent2017,henryImagebasedSexualAbuse2021} note that ``much of the harm that perpetrators are able to cause...is due to the social and cultural sexual taboos attached to''~\cite{henryImagebasedSexualAbuse2021} sex and sexual content. Sex-positive approaches have been recommended to help alleviate the stigma and trauma-related symptoms victim-survivors face~\cite{baggettSexpositiveAssessmentTreatment2017, ocallaghanThinkJustHaving2024}, and to guide the creation of sex education and policies that reduce victim-blaming and sexual violence perpetration~\cite{williamsResolvingSocialProblems2013, williams2015moving, peterson2024can, sower2023kink, carmody2005ethical}.

\subsection{Generative AI and Sexual Content}
\noindent\textbf{AIG-SC.} 
\new{Prior analysis of AIG-SC on Reddit demonstrates that a wide array of individuals are discussing the topic, including those who are interested in creating AIG-SC content for entertainment and those who are perpetrating AIG-NCC~\cite{doringExperiencesAIGeneratedPornography2025}. 
Content creators report having both positive and negative experiences associated with AIG-SC. While some reported experiencing pleasure or having fun, others (particularly AIG-NCII perpetrators) recounted feelings of shame and stress.}
Other research on AIG-SC has focused on monetized AIG-SC and its potential impacts.
As generative AI capabilities for sexual content advance, there is a burgeoning ecosystem of \new{individuals~\cite{dawoudUndergroundMainstreamMarketplaces2026}} and businesses seeking to profit~\cite{emanuelmaibergBoomingAIPimping2024}.
Anciaux and Gramaccia \cite{anciauxImaginingMarketsCrafting2025} demonstrate
that the current ecosystem primarily consists of three types of business ventures: those that sell AIG-SC directly (e.g., through subscription-based platforms or commissions), those that help sex workers automate tasks and scale content creation, and intermediary services targeted at other AIG-SC creators. \new{Dawoud et al.'s exploration of AIG-SC on Fiverr further confirms that commissions are rapidly growing. They also observed that sellers frequently advertise that their content can be used downstream on OnlyFans, Fanvue, and social media platforms, for potential use by those who are competing with human-made content on said platforms and also sex workers who are interested in using AI to scale content creation~\cite{dawoudUndergroundMainstreamMarketplaces2026}. AIG-SC sites that offer subscription-based content generation are also widely available and enable those without any specialized skills to create highly customized content through pre-defined feature selection (e.g., age, weight, clothing choices) and user-inputted prompts (e.g., including image uploads that potentially enable NCII)~\cite{lapointePresentFutureAdult2025}.}

Other work has explored queer artists' desires to use AI for artistic creation, finding that ``safety'' interventions restrict artistic expression and representation of queer identities~\cite{taylorUnStraighteningGenerativeAI2025}, extending broader concerns about algorithmic suppression of nudity in art~\cite{riccio2024exposed}.  
Alongside the growing AIG-SC economy, there are increasing concerns about impacts to the sex industry and the nonconsensual use of sex workers' content.
While content theft from sex workers is not new~\cite{zahrastardustSexTechAge2023, stardustSEXTECHINDUSTRIESCULTURES2025, mcdonald2021s}, AIG-SC increases the threat of labor displacement. 
Similar to other performing and creative fields~\cite{kyi2025governance, jiang2023ai, goetzeAIArtTheft2024}, academics emphasize the necessity of including sex workers in AIG-SC development to leverage their expertise and ensure their inclusion in profit~\cite{butlerSexWorkerHuman2025}.

\header{AIG-NCC}
\label{subsect:related-work-NCII}
\revision{Prior work on AIG-CSAM in the computing community has tested the efficacy of different AI safety strategies in preventing the generation of such content~\cite{baSurrogatepromptBypassingSafety2024, cretu2025evaluating}, the impact of handling such content on the people who test AI systems~\cite{pendseWhenTestingAI2025}, and the preparedness of teachers to handle AIG-CSAM in a K-12 school setting~\cite{wei2025utterlyillprepared}.} Work on AIG-NCII has examined its prevalence and harms~\cite{umbachPrevalenceImpactsImageBased2025,umbachNonconsensualSyntheticIntimate2024}\revision{, motivations for creating and sharing} AIG-NCII~\cite{flynnSexualizedDeepfakeAbuse2025} and attitudes towards it~\cite{brighamViolationMyBody2024, umbachNonconsensualSyntheticIntimate2024}.
Other works \revision{have analyzed the methods used to create AIG-NCII,} including nudification apps~\cite{gibsonAnalyzingAINudification2025},
marketplaces~\cite{hanCharacterizingMrDeepFakesSexual2025, timmermanStudyingOnlineDeepfake2023},
and model repositories~\cite{hawkinsDeepfakesDemandRise2025, weiExploringUseAbusive2024}.

\header{Regulating AIG-SC and AIG-NCC}
\facct{While the technology that enables AIG-NCC is becoming more accessible, global laws to criminalize its creation are still emerging~\cite{michelledonelanOnlineSafetyAct2023, tedcruzS146TAKEIT2025} and vary in approach:
the European Union is considering banning the marketing of AIG-NCC-enabled technology~\cite{pieterhaeckEUSetBan2026},
Denmark has begun allowing their citizens to hold copyright over their own likeness~\cite{Guardian_Denmark_Deepfakes_2025},
and South Korea has criminalized both the possession and viewing of AIG-NCC~\cite{Reuters_SouthKorea_Deepfakes_2024}.}
The U.S. TAKE IT DOWN Act~\cite{tedcruzS146TAKEIT2025} distinguishes AIG-SC from AIG-NCC using an identifiability standard: AIG-NCC depicts a recognizable real individual without that individual's consent (anyone under 18 cannot give consent). Yet, no technical specifications exist that operationally define this standard for a given piece of content depicting an AI-generated ``person.'' 
Even if separated from AIG-NCC, some AIG-SC content may be deemed obscene and thus illegal, but the remaining AIG-SC \facct{is protected in jurisdictions where viewing and distributing sexual content are legal.}
\facct{Definitions for obscenity are also contentious. For example,} the U.S. definition of obscenity is notoriously vague~\cite{karnielPornographyCommunityInternetFreedom2004}, with a long history of overbroad definitions~\cite{fullerPornWarsSerious2019} being abused for other purposes like criminalizing the distribution of lawful abortion-inducing medication~\cite{beckAntiAbortionExtremistsWant2024}. Even the legality of fictional explicit representations of minors such as drawings and cartoons that are not identifiable as a real child~\cite{18USC1466A2003,internetwatchfoundationHowAIBeing2023} is hotly debated~\cite{cretu2025evaluating, abuseStateLawsCriminalizing2025,kokolaki2025unveiling} and human raters exhibit significant disagreements in identifying children in AI-generated images~\cite{cretu2025evaluating, kireev2025manually}.
\facct{Similarly, outside of the U.S., 
imagery including clothed individuals holding hands could be considered intimate and violating if non-consensual~\cite{batoolExpandingConceptsNonConsensual2024}.}
Given these challenges, regulators and technical standards bodies offer few concrete recommendations for operationally defining a piece of content as protected vs. unprotected speech or for preventing AI systems from generating AIG-NCC or obscene AIG-SC~\cite{chandraReducingRisksPosed2024}. In this gray zone, unregulated third-party and open-source AIG-SC platforms have proliferated, with an increasing
reliance on inconsistent and ever-changing industry-based governance~\cite{milmoOpenAIConsidersAllowing2024}.

\header{AI Companion Bots}
\revision{Finally, AI companion bots have also received growing attention.
By allowing users to customize, converse, and engage in sexual roleplay with AI characters, sites such as Replika~\cite{lukainc.Replika2025} have gained significant popularity~\cite{hansonReplikaRemovingErotic2024}; however, it is unclear whether companion bots are beneficial to users.
While some works highlight how bot-use may reduce loneliness and assist in social skill development of users~\cite{guingrichChatbotsSocialCompanions2025}, others worry they reinforce unhealthy emotional dependence~\cite{laestadiusTooHumanNot2024} and conventional beauty standards which flatten women into stereotypes~\cite{depountiIdealTechnologiesIdeal2023}.}

%% file: arXiv-sections/methods.tex
\section{Methodology}
\label{sec:methods}
To gather insight into how AI is used in the production of sexual content,
we conducted $N$=$\numParticipants$ semi-structured interviews with a range of adults who generate sexual content with AI. See~\cite{authorsUnlimitedRealmExploration2025} for our recruitment material, interview questions, and codebook.

\subsection{Study Scope}
While we recruited participants using a variety of methods (see \textbf{Recruitment}), a majority of our participants were recruited from an online \AIGSC{} community. 
\revision{While prior work has observationally studied online communities that exist solely for creating and sharing AIG-NCC~\cite{hanCharacterizingMrDeepFakesSexual2025, timmermanStudyingOnlineDeepfake2023, weiExploringUseAbusive2024, hawkinsDeepfakesDemandRise2025}, the communities we contacted for recruitment purposes were focused on AIG-SC and had stated rules prohibiting the sharing of AIG-NCC.} \revision{Given that our goal was to understand the broader landscape of AIG-SC, we did not intentionally seek out those who create AIG-NCC. Still, a few participants disclosed this behavior to us during their interviews. It is also possible that other participants may create such content but did not disclose doing so.}

Participants often described AIG-SC creation as a ``passion,'' with several participants reporting spending many hours a day creating content. \revision{The characteristics of the average AIG-SC creator are not yet known; however, given our recruitment method, our participants may skew toward more active creators and thus they may invest} more time, effort, and skill development than others who create AIG-SC. \revision{As we interviewed active AIG-SC creators, many of whom are in a community with other creators, our participants likely have} positive perceptions of AIG-SC, and \revision{may be} more likely to argue in its favor. \revision{Since this study focused on motivations for creation, we deliberately centered the experiences of active creators. Future work should analyze the perspectives of non-creators, particularly sex workers and those who participate in online communities where sexual content, including (unwanted) AIG-SC, is shared~\cite{maibergRedditHentaiCommunity2024}.}

\subsection{Research Procedures}

\noindent
\textbf{Recruitment.} 
From April to August 2025, we recruited a sample of \numParticipants{} participants via convenience sampling and targeted recruitment.
First, we reached out to those within our networks to disseminate our recruitment flyer (see~\cite{authorsUnlimitedRealmExploration2025}).
We also identified a list of large online communities  
relevant to our study and reached out to moderators for consent in recruiting members of their community.
Lastly, we used snowball sampling by asking participants to refer us to others. Each participant provided informed consent and filled out a short screening survey on their background in AI and AIG-SC. Participants could voluntarily provide their demographics. We took background information and demographic information into account when selecting participants to ensure diverse perspectives. Participants were paid $\$40$ USD (or an equivalent amount in another currency) through a gift card or PayPal; additionally, participants were compensated $\$10$ USD for each referred participant who completed an interview.

Our 28 participants were from a wide range of demographic backgrounds\new{\footnote{Participants were asked optional free-form questions about their gender and sexual orientation. We categorized their free-form responses to their closest representation. Our presentation of summary statistics therefore does not represent the full richness in participants’ self-descriptions. Several participants identify as multiple genders, and thus this
count is not exclusive.}} (as fully detailed in Appendix~\ref{appendix:demographics}).
\new{Twenty identified as men, 
five as women, and four as non-binary. Eighteen identified as heterosexual and nine as bisexual, gay, lesbian, pansexual, or queer. One participant declined to answer optional questions on gender and sexual orientation.}
While participants' median age was between 25-34 years, they ranged from 18 to 55+ years.
\facct{Our participants were located across 13 countries including the United States ($n$=$11$), Canada ($n$=$4$), Mexico, France ($n$=$2$ each), Australia, Brazil, Chile, China, Colombia, India, Indonesia, the Netherlands, and the United Kingdom ($n$=$1$ each).}

\header{Semi-Structured Interview}
\new{
We asked each participant about (1) their background with sexual content creation and AI, (2) their creation process, (3) their interactions with AIG-SC-centric communities, and (4) any content creation or sharing boundaries they or others have.} \new{This work centers  participants' responses to the first three sections. The remaining data will be featured in future work.}
Interviews lasted 73 minutes on average, totaling 34 hours and 22 minutes. 
\new{The research team consists of two cisgender women and one non-binary, assigned male at birth, masculine-presenting person. 
Informed by prior work, which shows that matching the presented identity of interviewers and participants increases the disclosure of sexual violence and marginalized sexual behavior in academic studies~\cite{harling2019influence, wilsonEffectsInterviewerCharacteristics2002}, interviewees were interviewed by an interviewer with the same gender presentation when possible.
Interviews were conducted in English.}

\smallskip
\noindent
\textbf{Analysis.}
Our team followed an inductive coding process and conducted a \revision{reflexive} thematic analysis~\cite{braun2006using}.
Each author began the process by independently reviewing a selection of transcripts\footnote{\revision{To improve clarity, minor edits (such as the removal of filler words) were made during transcription~\cite{corden2006using}.}} to openly explore themes that emerged. 
The authors then independently created codebooks and met afterward to converge on an initial codebook. This codebook was applied to three transcripts that the authors triple-coded. 
The authors met to review and discuss any discrepancies and to surface any new codes that arose while coding. 
The authors then applied the codebook to two additional transcripts. We held one more review of our codes to finalize the codebook (see~\cite{authorsUnlimitedRealmExploration2025}). 
The remaining transcripts were divided among the three authors for independent coding.

\subsection{Terminology}
\label{subsect:terms}
Since many (but not all) participants were actively involved in online AIG-SC communities, we briefly explain the structure and terminology participants use when discussing them.
\new{As shown in Figure~\ref{fig:community-architecture}}, these communities often exist on a third-party digital service (a ``platform'').
These platforms provide digital infrastructure that lets groups establish public and private spaces (a ``community'') that people can join as ``members''. Privileged members (``moderators'') further structure the community and moderate other members (e.g., by enforcing stated rules or the platform's terms of service).
Within these communities, moderators may set up digital sub-spaces (``channels'') that contain message boards to post images and text, or synchronous voice/video channels for members to directly interact with each other. Multiple channels may be set up, each dedicated to specific interests or content.
\revision{When quoting participants sampled from these communities, we refer to AIG-SC creators as ``CXX'', and those who also disclosed perpetrating AIG-NCC as ``PXX''.}

\subsection{Limitations}
While our recruitment and interviews were in English, in some cases, English was not the participant's primary language. 
AIG-SC extends beyond English-speaking countries and persons, and additional work should be conducted to understand AIG-SC in other cultural contexts.
Furthermore, desirability bias could influence our results as participants were potentially less likely to engage in, or report engaging in \revision{stigmatized activities and AIG-NCC creation.}
\revision{As mentioned, our study focuses for the most part on the perspectives of highly motivated content creators who invest significant resources in AIG-SC, whether money, hardware, or time.
For example, \Pten{}'s daily routine includes generating AIG-SC:
\pquote{I wake up, I go into Civit.AI... I download maybe two hundred gigabytes... [I] generate maybe one or two, maximum three, images daily...
And when I have vacation... I [take] it more seriously to develop a more complex image using new techniques.}}

\subsection{Ethical Considerations}
\input{arXiv-sections/ethics}

%% file: arXiv-sections/ethics.tex
\label{sec:ethics}
In setting out to understand the who, what, and why of AIG-SC, the team engaged in continuous discussion throughout the project on how to conduct this research responsibly and ethically for the multitude of parties involved, including the research participants, the broader population of AIG-SC creators, those whom AIG-SC may impact (sex workers, artists, writers), victim-survivors of AIG-NCC, victim-survivors of intellectual theft whose data may be non-consensually used in model training, and the safety of the research team. This scope was not limited to the research procedures, but also considered the expected impact this work may have. To the best of our ability, we summarize the ethical discussions among the team, and the resulting procedures we followed.
All procedures were independently approved by the Institutional Review Board of each author's institution.

\header{Encounters with Ethical AIG-SC Creators}
We followed best practices in ensuring participant privacy for the AIG-SC creators we intended to recruit. Due to the stigma of creating sexual content, we added additional measures to further protect participant privacy, autonomy, and dignity. We only recruited from large online communities (>1000 members). Prior to recruitment, we received permission from the moderators before posting any messages or materials.  Participants were given the option to conduct their interview via text chat rather than video or audio, which may inherently be identifying. Six participants requested to be interviewed exclusively over text chat. 
We provided instructions on how to create an anonymous email account for correspondence regarding this study and instructions for changing Zoom usernames. Prior to each interview, the purpose of the study was re-disclosed, and each participant was reminded of their rights and re-affirmed consent. Participants were also given the option to withdraw their data or redact any text from the interview afterward, though none did.
After creating a full draft of our findings, we did selective member checking~\cite{mckim2023meaningful} by sharing the draft with each online community moderator to elicit feedback about the work's accuracy and identify any privacy leaks evident to someone who knew the community intimately; when appropriate, suggestions were incorporated.

\header{Potential Encounters with Perpetrators of Sexual Violence}  We recruited only from online communities that had stated rules prohibiting the sharing of AIG-NCC. Despite this, we prepared for the potential of encountering perpetrators of AIG-NCC among those we recruited. 
While it is well-established in psychology and criminology that research with sexual violence perpetrators is important towards reducing sexual violence (e.g., Fisher et al.~\cite{fisher2002adult} interviewing adult sex offenders to understand who, how, and why), in computing,
with the exception of two recent publications~\cite{umbach2025perpetration, jimenez2024understanding}, perpetrator behavior is infrequently studied through direct interaction.
This is for no small reason as research with perpetrators of sexual violence demands great care. However, these insights also provide key knowledge that may otherwise be unobtainable, yet essential in informing effective public policy and technical defenses (such as perpetrators' motivations, actions, and justifications)~\cite{cowburnConfidentialityPublicProtection2005}.
As the goal of this work is to inform governance of AIG-SC that is accurately situated in the existing reality of actors and actions, and community-led governance, or lack thereof, including governance against AIG-NCC, data collected from AIG-NCC perpetrators, recruited from general AIG-SC communities, is critical in providing an accurate picture of the context in which governance must take place. Such context can inform the shaping of community rules (as seen already by one interview participant prompted only by the interview itself) and design of interventions.

We followed established guidelines for interviewing sexual violence perpetrators~\cite{hearn2007background, sexual2012ethical, sikweyiya2011perceptions, cowburnConfidentialityPublicProtection2005}.
To prioritize the safety of the public and the research team, while also ensuring all participants were treated with rights afforded to them in the Helsinki Declaration~\cite{WMA2024Helsinki} (e.g., safe treatment, informed consent, minimization of risk, and adherence to an approved research protocol), we established clear, IRB-approved procedures for handling disclosures of sexual violence. Specifically, we prepared procedures for when the confidentiality of an interview must be breached, how interviewers should respond verbally to the disclosure or justification of NCII, and for researcher protection.

As we had prepared for, three of our participants, recruited through our general recruitment strategy for AIG-SC creators, turned out to be perpetrators of AIG-NCC. 
\new{None of these participants disclosed AIG-NCC creation in their screening surveys but through the course of the interview, they disclosed AIG-NCII perpetration when describing content they generated.} 
We now discuss our procedures for handling these disclosures.

\smallskip 
\noindent 
\textit{Limited Confidentiality:}
Following established guidelines for researching perpetrators of sexual violence~\cite{hearn2007background, sexual2012ethical, sikweyiya2011perceptions, cowburnConfidentialityPublicProtection2005}, we prioritize the current and future safety of victim-survivors. 
If, during the course of the interview, the participant disclosed a detailed account of themselves committing a criminal offense in which they create CSAM, publicly post AIG-NCC of an identifiable victim, or state clear immediate intentions of doing so, we would end the interview (if active)\footnote{In the case of \Pfortyfive{}, whose publicly posted NCII we reported, we were not certain that they had created visual media and publicly posted the content until we viewed their social media profile (which they provided to us) after the interview.} and report the activity to \revision{NCMEC CyberTip line~\cite{nationalcenterformissing&exploitedchildrenNationalCenterMissing2026} (in the case of CSAM) and the platform authority (if the AIG-NCC was published online)}.
\facct{This limited confidentiality was clearly stated when obtaining consent from participants.\footnote{\facct{Specifically, our consent form notes~\cite{authorsUnlimitedRealmExploration2025}: ``In the event that you disclose illegal behavior, such as using generative AI tools to create sexual/explicit images of minors, we may be required to disclose this to appropriate authorities, which will result in a loss of confidentiality''. This was also verbally explained to participants before their interview, along with their other rights listed in the consent form.}}}
Following established practices and thought~\cite{cowburnConfidentialityPublicProtection2005}, as researchers, we did not act in a law enforcement capacity, did not attempt to elicit details that would breach confidentiality, and minimized the amount of information collected to achieve the goals of the research.
Ultimately, one participant sent researchers a public account in which NCII of an identifiable public figure was posted. After their interview, the research team discovered this NCII and reported the content.
After our study, community moderators were informed of the research results, including the existence of AIG-NCC perpetrators that may have been recruited from their community. In our future work, we intend to use these results to support ethical and effective governance of AIG-SC communities that respect the rights of free expression and privacy, without tolerance for abuse.

\smallskip 
\noindent 
\textit{Preventing Collusion and Condonement:}
We do not seek to normalize or justify any action of sexual abuse or perpetration of sexual violence, including the creation of AIG-NCC. 
Following established guidelines~\cite{sexual2012ethical, hearn2007background}, at no point in our interviews did the interviewer condone perpetrators' actions, or minimize harm of sexually violent actions, either verbally or non-verbally.
To prevent bias and possible confrontation, interviewers did not provide opinions on perpetrator behavior, but did question the acceptability of violent behavior if minimized during the interview~\cite{sexual2012ethical}.
Our interview seemed to trigger an awareness that AIG-NCC creation is unacceptable. 
In one case, \Pfortysix{}, who disclosed AIG-NCII perpetration when responding to an initial set of questions on their personal boundaries, later recanted having created AIG-NCII or any sexual content. 
In a second case, on the same day of their interview, a community moderator added significant updates to their community rules that expanded boundaries against AIG-NCC.

\smallskip 
\noindent 
\textit{Researcher Protection:}
Given the potential exposure to AIG-NCC perpetrators, we protected researcher safety by consulting with researchers who previously conducted perpetrator interviews, including those from non-CS fields that have established norms and a history of engagement in research on perpetrators of sexual violence, and openly discussed concerns and mitigations as a team\facct{. These mitigations included (1) matching interviewer and interviewee gender presentation, as AIG-NCC often occurs across gender lines; 
(2) minimizing researcher identities beyond required disclosures (consent forms); 
and (3) preemptively removing online information of the researchers.}
Realizing that such a study inherently carries non-negligible risk to the researchers (e.g., researchers may become the target of AIG-NCC), and that this was the team's first work with these communities, we aimed to mitigate power dynamics by not including any persons in degree-granting programs (e.g., PhDs, Masters, Undergraduates). 
Our team was composed of researchers who had completed their training and opted in with full autonomy. While we believe in the value of including students in research on this topic, we made this decision out of an abundance of caution, given that we were interacting with AIG-SC creators for the first time. 
Through this process, we have built relationships with moderators within AIG-SC online communities and moving forward, we are conducting follow-up work in teams that include students. Researchers had access to a mental health consultation at any time.

\smallskip 
\noindent 
\textit{Academic Reporting on AIG-SC Methods:} \facct{As academic publications have been leveraged by AIG-NCC creators~\cite{hanCharacterizingMrDeepFakesSexual2025}, we also consider the impact our own publication may cause.
In this publication, we do our best to ensure our findings are not framed in an instructional way or introduce harm. Unlike work introducing new AIG capabilities, we document and report on methods already widely used in AIG-SC communities. 
We acknowledge that this inherently provides greater visibility into these methods, but we also believe it is necessary to inform academic and policy audiences of ongoing harms to address and remediate them.}

%% file: arXiv-sections/background-and-content.tex
\section{\textit{Who} is Creating AIG-SC? (\ref{rq:background})}
\label{sec:AIG-SC-Content}
\input{tables/participants}

\revision{
}
\label{sec:creator-backgrounds} 
Similar to creators who use AI for general art~\cite{tangExploringImpactAIgenerated2024}, some AIG-SC creators had relevant professional backgrounds, while others did not (see Table~\ref{tab:part-expr}). \revision{\Pthirtytwo{}} moderates a large AIG-SC community with:
    \begin{quote}
    \pquote{3 categories of people... (1) people who are already pretty experienced with AI and just looking for a place to be with others like them,
    (2) people who have a bit of experience using ChatGPT but don't know anything else and need guidance on how to use other models,
    (3) people who know absolutely nothing, they have heard they can generate AI porn and come here to learn how.}
        \end{quote}

Specifically, while only three participants had formal education in \revision{AI}, 11 participants had prior experience coding and eight were employed in the tech industry; two participants were involved in black-hat communities. 
A few participants were professionally employed in `creative industries' as videographers (\Pfour{}), art contractors for enterprises~(\Pten{}), commission takers~(\Pthirtyseven, \Ptwentynine), and independent, exhibited artists~(\Ptwentysix{}); all creative industry participants created sexual content as part of their creative work. 
Finally, three participants had worked in the sex industry: two participants were sex workers, and the aforementioned videographer was an experienced pornography videographer.

\header{Introduction to AIG-SC} 
Participants were first introduced to the idea of using AI to create sexual content in a number of ways. For some, it was an immediate thought after seeing AI-generated images: 
\pquote{One day I was like, `What if I just type in ``naked woman''?' [chuckles] And it was like, `Yo! It just worked.'}~(\Peleven{}). Others saw AI as a solution to an existing gap in the availability of sexual content that met their needs: \pquote{Erotica that other people made are not hitting that right spot, so when I hear[d] that GPT can write erotic story (from Reddit I believe) I start[ed] trying it out}~(\Pnineteen{}).
Others used AI for non-sexual content but encountered platform safety filters that drove their curiosity --~\pquote{[When] using it to generate stories and poetry... I got very annoyed at the sanitization of ChatGPT from filters, so I searched for NSFW AI communities}~(\Pthirtysix{}) -- or found \revision{that the AIG-SC community was more knowledgeable and supportive while learning AI than other online spaces:}
\new{\pquote{Porn has been the thing that advances [technology]... some of the best VR technology is through porn, and it's the same thing with AI... %this is the same with AI image generation...
[AIG-SC creators] will spend all day [generating images], because they love to do it... 
[Compared to non-sexual spaces], they actually are the people that work harder, and that’s why it's better to work with these people}~(\Ptwelve{}).}

%% file: tables/participants.tex
\begin{table*}[t]

\centering
{
\small
\definecolor{rowgray1}{gray}{0.97} 
\definecolor{rowgray2}{gray}{0.92} 
\rowcolors{2}{rowgray1}{rowgray2}

\makebox[1.00\textwidth][c]{
\begin{tabular}{lllll}
\toprule
{\bf ID} & {\bf Background} & {\bf Use Case} & \multicolumn{2}{c}{\bf Technical Set-Up} \\
\cmidrule(lr){4-5}
 & & & {\bf Text} & {\bf Image}  \\
\midrule
\midrule
\Pone & Sex Work & Scaling image content (depicting self), writing erotica &     
\ding{51}(\textbf{S})$\rightarrow$\ding{51}(\textbf{M}) & \ding{51}(\textbf{S})

\\
\Pseven & Sex Work & Scaling image content (depicting self), 
scaling messages & 
\ding{51}(\textbf{S}) & \ding{51}(\textbf{S})
\\
\Pfour & Sex Industry/Videography & Assisting with creating pornographic films & \ding{51}(\textbf{M}) & 
\ding{51}(\textbf{M}/\textbf{S})

\\
\Pthree & Software/Sales & Building a commercialized AIG-SC platform

& 
\ding{55} & \ding{72}(\textbf{L}/\textbf{C})
\\
\Pthirtyone & Tech; Hobbyist Writer & Creating anime-style images, writing erotica & \ding{51} & \ding{72}(\textbf{L})
\\
\Pthirtytwo & N/A & Roleplaying, creating jailbreak guides & 
\ding{51}(\textbf{S})$\rightarrow$\ding{72}(\textbf{L})$\rightarrow$\ding{51}(\textbf{M}) & \ding{55} 
\\
\Pfortyfive & Language Teacher & Professional AI red teamer; generated AIG-NCC & \ding{51}(\textbf{M}) & \ding{51}(\textbf{M})

\\
\Psixteen & IT Professional & Roleplaying & \ding{51}(\textbf{S})$\rightarrow$\ding{51}(\textbf{M}) & \ding{55}
 
\\
\Pseventeen & Content Consumer & Roleplaying & 
\ding{51}(\textbf{S})$\rightarrow$\ding{51}(\textbf{M}) & \ding{55}
\\
\Pnineteen & Hobbyist Photography & Roleplaying  & 
\ding{51}(\textbf{M}) & \ding{55} 
\\
\Pthirtysix & Writer/Artist & Roleplaying, creating fandom-based images & 
\ding{51}(\textbf{S})$\rightarrow$\ding{51}(\textbf{M}) & \ding{72}(\textbf{C})
\\
\Pthirtyseven & Semi-Pro Artist & Share likeness LoRAs, art reference, roleplay, audio & \ding{51}(\textbf{M}) & \ding{115}/\ding{72}(\textbf{C}) 
\\
\Pthirtynine & Software Dev. & Creating jailbreaks for others & \ding{51}(\textbf{M}) & \ding{51}(\textbf{M})

\\
\Pfortytwo & Filmmaker/Writer & Jailbreaks, prompt dev., erotica, creative writing & \ding{51}(\textbf{M}) & \ding{55} 
\\
\Pfortythree & Professional Artist & Roleplaying & 
\ding{51}(\textbf{S})$\rightarrow$\ding{72}(\textbf{L})$\rightarrow$\ding{51}(\textbf{M}) & \ding{55}

\\
\Pfortyseven & Gaming/Writing Hobbyist & Roleplaying & \ding{72}(\textbf{C}) & \ding{72}(\textbf{C})

\\

\Pfive & Software Dev. & AIG-SC for art assets in video game & 
\ding{72}(\textbf{C}) & \ding{72}(\textbf{C})
\\
\Ptwentyseven & Multimedia Design & Roleplaying, AIG-SC for art assets in video game & \ding{72}(\textbf{L}) & \ding{72}(\textbf{L})
\\
\Ptwo & AI/STEM & Building a commercialized AIG-SC platform & 
\ding{55} & \ding{72}(\textbf{C})
\\
\Ptwelve & YouTuber/Marketing & Generating character-based images  & 
\ding{55} & \ding{72}(\textbf{C}) 
\\
\Ptwentysix & Professional Artist & Generating kink-specific images & 
\ding{55}  & \ding{115}/\ding{72}(\textbf{L}) 
\\
\Ptwentynine & Light Content Creation & Creating images for followers/commissions & \ding{55} & \ding{115}/\ding{72}(\textbf{L})
\\
\Pfortyone & Disabled Artist & Used AI to create (sexual) art independently & \ding{55} & \ding{51}(\textbf{S})
\\
\Pfortysix & CS/Graphic Design & Creating images, including AIG-NCC 
& \ding{55} & \ding{51}(\textbf{S})/\ding{72}(\textbf{C})
\\
\Peleven & Freelance Software & Creating images, including AIG-NCC (e.g., fake OnlyFans) & \ding{55} & \ding{115}/\ding{72}(\textbf{L}) 

\\
\Pthirtyfour & CS\&EE/Digital Pirate & Creating content of marginalized sexual orientation  & \ding{55} & \ding{72}(\textbf{L}) 
\\
\Pnine & Software Dev. & Training NSFW LoRAs for entertainment and challenge & 
\ding{55} & \ding{72}(\textbf{L}) 
\\
\Pten & Tech Artist & Creating images for entertainment and challenge & 
\ding{55} & \ding{72}(\textbf{L}) 
\\
\bottomrule
\end{tabular}
}

\caption{\textbf{Participants' Experience} --  \textmd{{\small 
Our sample contained a diverse set of $N$=$\numParticipants$ participants.
Here, we present their use cases and brief descriptions of their technical setups (see Section~\ref{sec:creator-backgrounds}).
Setup Symbols:  
\ding{55} = Wasn’t generated or unclear;  
\ding{51} = Pre-built System (\textbf{S}pecialized or \textbf{M}isused);  
\ding{72} = Customized Pipeline (\textbf{L}ocally- or \textbf{C}loud-hosted);
\ding{115} = Sold Commissioned AIG-SC Content. An ``$\rightarrow$'' denotes an evolution in their primary setup.
}}
}\label{tab:part-expr}
}

\vspace{-0.18in}
\end{table*}

%% file: arXiv-sections/use-cases.tex
\section{\textit{What} AIG-SC is Created and \textit{How}? (\ref{rq:content})}

\label{sec:what}
\revision{AIG-SC content is primarily either visual (e.g., images/video) or textual. 
Visual content is often made to depict specific personas, kinks, and styles, while text content typically allows AIG-SC creators to participate in roleplaying scenarios with a variety of personas and contexts.
While pre-built AIG-SC-specific applications were mentioned by some participants, several described building their own pipelines using open-source models or discovering jailbreaks in enterprise systems to create AIG-SC content.}

\subsection{Visual Content}
\revision{Most} of our participants \revision{($n$=$22$)} focused on generating visual content. 
While some participants have attempted to create videos \revision{($n$=$9$)}, it was generally noted that it was still costly and the results were low-quality when compared to static images:~\pquote{Photo quality it’s almost there but the video’s a long ways away}~(\Pfour{}).\footnote{\revisionthree{In less than a year since these interviews were conducted, \Pthirtytwo{} reached out to comment that \pquote{between the time this quote was taken and now, things have changed, and now video is really getting there... things are moving so fast that something you said yesterday is already outdated.}
}}
Participants generally attempted to generate content of the following forms (specifics are detailed in Section~\ref{sec:usecases}): 
\begin{enumerate}
    \item \textit{Specific Fictional Persona:}    Artificial human personas or fictitious characters and sentient creatures (either new or existing cartoon or anime characters) consistently represented over multiple AI-generated outputs.    \item \textit{Specific Kinks and Aesthetics:} Personas were frequently  depicted in combination with specific kinks related to poses and situations, and/or specific aesthetics around gender presentation or sexual orientation (see Section~\ref{sec:usecases:sexualexpression}).
    \item \textit{Specific Styles:} Often, creators generated images of a consistent visual style. Hyper-realistic and animated cartoon styles were most common, but also scenes that appeared cinematic or which fit within a specific style associated with the character depicted were noted.
\end{enumerate}

To generate images, participants mentioned three common methods.
\begin{enumerate}
    \item \textit{Custom AI Pipelines:} 
Many participants \revision{($n$=$16$)} built, trained, and generated images through a custom pipeline. These may consist of open-source models (e.g., Stable Diffusion~\cite{podell2023sdxl}), input prompts (e.g., Danbooru Tags~\cite{danbooru2021}),\footnote{\revision{An anime-specific tagging system for features within photos, such as characters or styles; these tags assist in image training and generation~\cite{danbooru2021}.}} and adapters (e.g., Low-Rank Adaptations~\cite{huLoraLowrankAdaptation2022} or ``LoRAs'')\footnote{\revision{A low-cost method of model retraining~\cite{huLoraLowrankAdaptation2022} that allows for specific styles or concepts in generated images~\cite{yang2024low}.}} which could be set up locally or on the cloud (e.g., PixAI~\cite{pixaiPixAIWebsite}).

    \item \textit{Pre-Built AI Applications:} \revision{Some} participants \revision{($n$=$7$)} also mentioned using simple, UI-bound~\cite{freedStalkersParadiseHow2018} applications to create AIG-SC. They either \textit{misused proprietary} AI applications (e.g., jailbreaking DALL-E (\Pfortyfive{}); extended sexual videos via Photoshop (\Pfour{}) 
    or leveraged \textit{specialized} AI applications such as nudification apps~\cite{gibsonAnalyzingAINudification2025}). 
    
    \item \textit{Commissions:}  \revision{A few participants ($n$=$4$) reported taking commissions for creating custom image content.
    Those who received commissions often shared content they had created and had an existing following on social media or within an AIG-SC community. They may take requests to continue growing their following, for monetary purposes, or for their own enjoyment.}
    Commissions were typically initiated through direct messages on servers and art boards, requests posted online, or existing gigwork sites; several participants reported being solicited for or working to create image AIG-SC for a client:~\pquote{The hobbyists will sell [custom] content... They’ll have Patreon, they’ll say, `Hey, I can make you any content that you want, reach out to me'}~(\Ptwo{}).
 
\end{enumerate}

\subsection{Text Content}
\revision{Many participants ($n$=$17$) generated text-based content and} engaged in two activities with text-generating AI systems: 
\begin{enumerate*}[nosep]
\item Creating and reading stories with sexual content; and
\item Roleplaying different narrative scenarios with sexual content. 
\end{enumerate*}

To generate this content, participants mentioned two methods:
\begin{enumerate}
    \item \textit{Custom AI Pipelines:}
    \revision{A few participants ($n$=$5$) created custom text-generation pipelines.}
    Roleplayers\revision{, in particular,} tended to have complex prompting pipelines for generating AIG-SC:~\pquote{It’s not just a blank prompt that you enter stuff into... It saves chats...    it lets you upload files}~(\Psixteen{}).
    Using a front-end like SillyTavern~\cite{coheeSillyTavern2025}, these structured prompts consisted of    ``character cards'' that describe the attributes of various personas in the narrative,
    the scene taking place,
    character dialogue,
    and jailbreaks (if necessary).
    While these prompts could be sent to open-source, locally deployed models, participants preferred using a proprietary model that they jailbroke.

    \item \textit{Pre-Built AI Applications:} 
    \revision{Alternatively, many participants \revision{($n$=$14$)} used existing, text-based user interfaces. Many participants} \textit{misused proprietary} AI applications, \revision{as} jailbreaking text models was seen as doable even by multiple participants with no AI or CS background.
    While systems like Claude were considered harder to jailbreak, systems like DeepSeek had few restrictions: \pquote{I’ve never even had to use a jailbreak for DeepSeek}~(\Pfortyseven{}). 
    Furthermore, participants mentioned that official policies are becoming more lenient:~\pquote {The new guidelines say that you’re allowed to write erotica now... So, effectively, ChatGPT is not safe for work}~(\Pone{}). 
    Participants also mentioned \textit{specialized} AI applications; a few were integrated into a larger application or experience (e.g., scaling messages to clients on adult content platforms~(\Pseven{}) and free-form video game dialog~(\Pfive, \Ptwentyseven)) but most were focused on chatting with character-based chatbots (\Pthirtytwo, \Psixteen, \Pseventeen, \Pthirtysix, \Pfortythree).

\end{enumerate}
While we did not hear reports of text being commissioned (perhaps because the content creation barriers are relatively low), \revision{tools for creating AIG-SC text were shared and sold (see Section~\ref{sec:usecases:financial})}.

\subsection{Jailbreaks}
Several participants discovered jailbreaks (\revision{i.e., methods for circumventing LLM restrictions~\cite{liu2023jailbreaking}}) to misuse proprietary models and create AIG-SC, while several other participants created jailbreaks without a direct goal of creating sexual content.
   \Pfortyfive{} professionally red teamed models and did so by creating content far beyond the policies companies attempt to enforce:~\pquote{I don't like reading this kind of text. I just find it easy to recognize: racism, hate speech, violent crime, chemical formula.}; unfortunately, this same rationale is used to perpetuate AIG-NCC (Section~\ref{sec:AIG-NCC}).   Others generated jailbreaks to help enable others to create the content they desire, such as \Pthirtynine{}, who regularly created and shared jailbreaks they discovered:~\pquote{I just kind of go toward whatever [AI model] people are having the trouble with the most.}    \revision{In such cases, understanding and discovering the jailbreak was the primary goal rather than the content it generates.}
    \Pfortytwo{}, who contributed in both the jailbreaking and AIG-SC creation communities, explained that while these communities are harmonious and cross over, AIG-SC creators and jailbreakers are often distinct:~\pquote{There are the jailbreakers... some of those people overlap and make NSFW stuff if they wish but for the most part the NSFW sphere is kind of different and distinct.}

\subsection{Other Content \revision{Modalities}}
While text- and image-based sexual content were predominantly reported by participants, other media were occasionally discussed.
\revision{A few participants ($n$=$4$)} created AIG-SC audio including erotic literature readings (\Pone{}, \Pthirtyfour{}), sensual sounds (\Pthirtyfour{}), and even sexually suggestive parody music (\Pthirtyseven{}): \pquote{If you've heard of Steel Panther, they're essentially 80s glam metal without the [subtle] innuendos, just straight up talking about what happens. I'll generate a lot of that type of content, parody content of different songs.}

    While most content was generated for just one modality, a few participants discussed generating multi-modal AIG-SC.
    For roleplaying, some participants used AI to visualize characters:
        \pquote{Let’s say... you wanted to do a story about... a specific demon. You have an idea what you want this demon to look like, so you use the AI image generation... it gives people a visual to work with}~(\Pseventeen{})
    
    Some participants discussed future ambitions that combine video and audio to make full adult films:~\pquote{I would love to direct the first fully AI naughty movie}~(\Pfour{}).
    Others note that these media are often used to create virtual AI companions that text and send images and video, colloquially known as an ``AI girlfriend/boyfriend'':~\pquote{[There's] something called the Eva AI, which creates AI digital copies of us [sex workers]... It’s all the digital girlfriend experience}~(\Pseven{}).

\section{\textit{Why} is AIG-SC Being Created? (\ref{rq:motivations})}
\label{sec:usecases}
\label{sec:motivations}

\revision{Many were motivated to create sexual content for sexual satisfaction. Some participants who desire more niche content replaced their other sexual content consumption with AIG-SC: 
\pquote{Nine times out of 10... I tend to generate an AI image over searching [for it online]}~(\Pthirtyseven). For others, AIG-SC diversified the sexual content they consumed by \pquote{mix[ing] in the real with the fake. Just like I'm not a vegetarian, but... sometimes when I would eat like a street dog, I would just get a vegetarian dog... even though I knew it was faux. I would just add a little bit of zest}~(\Pfortyone).}
Participants also created AIG-SC to express themselves artistically and creatively, \revision{to be in and build} community, \revision{explore} technical interests, or \revision{act on} beliefs they hold about AI censorship and safety. \revision{Others had (sometimes simultaneous) financial motivations for content creation: \pquote{some are here to make money, some are here to enjoy free time} (\Pfortysix).}

\revision{We emphasize that this section captures participants' self-described use cases and motivations, and thus the narratives are largely positive. \new{Their (limited) reflections on the potential negative impacts of AIG-SC are offered in Section~\ref{subsec:impacts}, including discussion on labor displacement~\cite{anciauxImaginingMarketsCrafting2025}, the reproduction of harmful sexual/gendered  tropes~\cite{depountiIdealTechnologiesIdeal2023}, and the spread of unwanted AIG-SC to other online communities~\cite{maibergRedditHentaiCommunity2024}.} We also note that some AIG-NCII perpetrators may justify their actions by claiming creative and artistic expression~\cite{flynnSexualizedDeepfakeAbuse2025}. We intentionally include and distinguish narratives held by creators and perpetrators in this section (by referring to creators as ``CXX'' and those who disclosed perpetrating AIG-NCII as ``PXX'' as mentioned in Section~\ref{subsect:terms}), which highlights the complexity of AI governance in which stated motivations and virtual space are shared.}

\subsection{Sexual Satisfaction and Exploration}
\label{sec:usecases:sexualexpression}
Participants reported that AI content creation enabled new forms of sexual exploration by granting more freedom and agency over the content they could consume: 
\pquote{You can’t change what the [porn] video does... but the amount of agency you have with a roleplay with an LLM is unparalleled...
I haven’t really looked at that much porn at all since I started doing this}~(\Pfortyseven{}).
Participants stated that AI allowed them to generate content to satisfy their niche interests, such as erotic stories involving unpopular characters within a fandom (\Pthirtysix{}) or a specific sexual kink, 
as described by \Ptwentysix{}: 
\begin{quote}
\pquote{It takes months to [find content] that is exactly what I like... [AI] really helped me just visualize like a lot of my specific kinks... made me like feel a lot of freedom in terms of being able to do anything I want with AI. And I don’t have to like wait for someone else to draw it or pay someone else to draw it.}\end{quote}

AI was also described as a way to explore different sexual situations and fantasies, including themes that are more difficult for human models to portray, such as \pquote{medieval settings...
fantasy, and sci-fi}~(\Pfortyseven{}). 
Particularly for those who engaged in roleplaying, they could more directly embody and engage with specific scenarios:
\begin{quote}
    \pquote{The beauty of using AI... is the unlimited realm of exploration and experimentation. You can interact with anyone. You can be anything... the sky’s the limit... 
    [You can be] a power dom. And the next one you’re the sub.. you can play a different gender... you could play a different species. You can literally do anything}~(\Pseventeen{}).
\end{quote}

This was particularly relevant for those who held alternative identities, allowing them to more fully embody them. For instance, \Pthirtyseven{} described how AI allowed them to explore their furry identity (their fursona~\cite{austin2023identity}): \pquote{I feel like [AI] has given furries a way to express themselves... I see a lot of people on CivitAI that make furry art, making their fursona for the very first time [by using] AI.}
AI was described as offering a few the opportunity to explore their gender and/or sexuality: 
\pquote{I've been questioning my gender identity for a while... [AI] did help me roleplay situations where I could act as a girl in everyday life situations, and then of course in NSFW situations}~(\Pthirtytwo{}). Others challenged norms of gender presentation: 
\begin{quote} \pquote{The character I designed [is] a female warrior... everyone told me maybe she’s missing something... she’s not strong [muscular] enough, because I couldn’t draw her strong without becoming ugly... I start experimenting with AI... to make her more like the idea I had of her... and start breaking those barriers about, `women need to be very slim'}~(\Pten{}).
\end{quote}

Participants generally held conflicting opinions about openness towards diverse content in online spaces related to AIG-SC but one participant found more acceptance toward content depicting trans people, which may suggest that AI can help fill gaps in the creation and sharing of more gender-diverse content for some~\cite{taylorUnStraighteningGenerativeAI2025}: \pquote{...a lot of Trans/Futa related stuff online in NSFW spaces is heavily fetishised and can be degrading, mainly from a male gaze...
its much easier with AI and in certain AI spaces}~(\Pthirtyone{}). 

AI may also afford people more privacy to explore their sexual interests, \revision{especially when fully local AI setups are used to create content on private devices (see Section~\ref{sec:what})}. This particularly affects people who are interested in exploring more stigmatized sexual interests or live in areas with stricter laws regulating sexual content: 
\pquote{You can just run it on your local hardware... for a lot of [my friends], it’s their number one concern. It is the privacy aspect}~(\Ptwelve{}). Unfortunately, this can also cause tensions in harm prevention.

Aside from content-based motivations, participants used AI to reduce logistics when roleplaying. This included no longer needing to identify someone with similar sexual interests. As \Psixteen{} expressed, \pquote{People very often write better than AI... but people have their own focus, their own lives, their own interests, their own preferences... [LLMs] are reliable...
They’re always there, they’re consistent.} 
Similarly, \Pnineteen{} wanted to avoid the complications of managing others' needs when roleplaying and more freely explore: \pquote{I'm just kinda bad at doing sexting... I don't wanna make other people feel bad because I don't make them feel good and vice versa. Because AI is not sentient, I don't need to care about that.}

\subsection{Creative Expression}
\label{sec:usecases:artistic}

Nudity and sexual expression have always played a role in art~\cite{riccio2024exposed}. Naturally then, for some participants who already had artistic pursuits or interests, AI is another medium. These participants considered their content creation to be an artistic endeavor, such as \Pthirtyone{} who shared that \pquote{I'm asexual but do find the human form exciting to express... [I've] always found erotic art to be the most expressive from all art I've seen.}
\Pfortyone{} likewise felt that the AIG-SC they created was more artistic than sexual. In addition to creating art, they were also involved in advocacy around sexuality and disability: \pquote{We were not shy of using art and [even] some explicit images to get some of our points across.} As a disabled artist, AI provided \Pfortyone{} with greater independence and autonomy in their art.
Previously, they were reliant on volunteers to assist them in bringing their visions to life. By using AI, they were now able to \pquote{do the work on their own} and explore sexuality in their work.

AI-generated content may be used as a reference for other artistic media, \pquote{so I can get a better understanding of how I can pose anatomical parts and also for coloring reference}~(\Pthirtysix{}) and because \pquote{with AI I can get it to generate almost exactly what my mind is picturing... AI can help envision what that would look like and give me ideas}~(\Pthirtyseven{}).

People also created content to showcase it online as AI-generated sexual art is itself an emerging genre: \pquote{I’m mostly drawing for myself and then going to exhibit it... 
in online galleries... right now I’m working on an artwork that’s just more like just a pin-up. Like a generic woman, but she’s like really fashion}~(\Ptwentysix{}). 

Some enjoyed creating AIG-SC purely as a creative outlet, \pquote{I get... just like a creative release. And um, like that's really healthy... like a really positive thing for me, to be able to like express myself even if it's like into the void and like, not like shared with anybody else}~(\Pfortyseven{}).

Beyond visual art, participants also noted they might use AI to help write and explore literary narratives, providing more creative control over narrative exploration. \Pseventeen{}~explains: \begin{quote}\pquote{If I read it in a book, then I’d be exploring somebody else’s creative vision... it wouldn’t have the same personal aspect because when I do these roleplays, I provide one of the characters. I am half of the equation... This AI [roleplay] stuff can absolutely be used for pure smut with no story... 
But in my personal taste, I prefer there to be a narrative...} \end{quote}
\revision{Several participants noted that they generally preferred reading and writing literature that contains darker themes, and they sought the same when roleplaying with AI:}
\pquote{[Other literature I've read] has sex, it has rape, it has hope, it has despair... it wouldn’t be what it is if it didn’t address the darkness... with an unrestrained AI, you can do the same}~(\Pseventeen{}). 
\revision{These participants felt that guardrails in large language models prevent narratives they wished to explore, echoing findings from prior work on queer artists' engagement with AI~\cite{taylorUnStraighteningGenerativeAI2025}. For example, \Pfortythree{} desired removing}
\pquote{the barrier that [AI] can't be lewd, that they can't make a flirty comment, that they can't talk about hard themes.}

\subsection{Technical Interest}
\label{sec:technical-interest}
Beyond content-related motivations, participants were drawn to the technical challenge of content creation: 
\pquote{I'll be kind of like ready to, you know, jerk off or something...
 but then it's like, I get like sidetracked by like the technical stuff... [I'll] forget that I was even into it on that sexual level in the first place}~(\Pfortyseven{}).
As discussed in Section~\ref{sec:creator-backgrounds}, many participants had technical backgrounds in computer science, software engineering, or machine learning. AIG-SC provided an outlet to \pquote{flex these skills}~(\Pthirtyfour{}).

Meanwhile, others developed technical skills through AI sexual content creation: 
\pquote{I've actually learned a bit of Python code throughout this. I learned how to use Git through GitHub and how to download certain programs}~(\Pthirtyseven{}).
Those who wanted to gain technical skills found that generating AIG-SC provided enjoyment while doing so:~\pquote{I wanted to learn the new [AI] tools and so I thought it was kind of a interesting creative outlet that was also stimulating as well}~(\Pthirtyfour{}).

Participants also viewed themselves as contributing technical knowledge to AIG-SC communities or innovating on the state of the art in generative AI: \textit{``It was less about satisfying myself and more about trying to generate the most realistic looking image''}~(\Peleven{}). 
This might entail focusing on certain features, such as 
\pquote{...perfect hands, perfect eyes}~(\Pfive{}).
Often participants described the novelty of creating content in this space:
\pquote{There’s this kind of frontier feeling where people are helping each other out to try and, you know, make something cutting edge, in that sort of spirit of like open source}~(\Pthirtyfour{}). This fueled desires to create content since it provided a challenge: \pquote{Because for me, it’s like this quest, or this test. Like, I want to see if I can make it}~(\Pten{}).

Relatedly, participants discussed an interest in boundary pushing to sate their curiosity in how far they could push AI. This included not only sexual content but also violent images. For example, \Peleven{}~\pquote{trained a model on the most violent images on the internet and it would just generate these horrible, cartel beheading pictures... I definitely enjoy trying to make something that someone else has had trouble making...
And that just so happens to be a lot of [chuckles] Not Safe For Work stuff. And if you can nail that, then you can nail anything.}
The inherent unpredictability of content outputted by AI models also added to the challenge and excitement of content creation: \pquote{There’s like also a surprise element, because it’s like you’re wrestling with this tool that doesn’t do exactly what you want...
So, it has that sort of, you know, variety randomness factor too}~(\Pthirtyfour{}).
This led \Pthirtyfour{} to develop a fixation, such that 
\pquote{there was a time when I was generating stuff... every day, just during my spare time, walking from one place to another... it has that sort of slot machine type of thing.}

\subsection{Financial Interest}
\label{sec:usecases:financial}

\revision{Some participants were interested in monetizing AIG-SC content (e.g., images, videos, stories) or assets along the creation pipeline (e.g., datasets, information on content creation).} 
Since only a few (\Pone{}, \Pfour{}, \Pseven{}) had a background in the sex industry (Section~\ref{sec:creator-backgrounds}), the vast majority \pquote{didn’t start monetizing [sexual content] until I started doing not safe for work AI}~(\Ptwentynine{}). This was primarily a ``side gig'' or a way to offset the costs of content creation: 
\pquote{AI is not free to run}~
(\Pseventeen{}).

Participants who sold content primarily did so via subscription-based platforms or direct commissions (e.g., through Patreon, DeviantArt, Fiverr).
 \revision{In both cases, those who are successfully monetizing content have built followings through social media and/or image sharing platforms.} 
Direct commission requests are often based on the quality of their work as demonstrated by public posts, their artistic style, or their offering of niche content.

\revision{
Creators also monetize their knowledge and other assets that are useful for content generation. 
\revisionthree{\Pthree{} observed that others were selling guides on how to create AI models for image generation, and \Pthirtytwo{} saw others selling guides for jailbreaking LLMs to produce NSFW content.} 
Others observed that people had sold specialized datasets (e.g., for a specific body part or kink), 
\pquote{[chuckles] like 'Middle-aged feet.' Something really specific that someone made and they’re like, `Oh, I got this now. Why not sell it in case someone else needs it?'}~(\Peleven). 
}

\revision{While some participants found opportunities for monetization after creating content as a hobby, others, who we describe as entrepreneurs, sought out the AI sexual content creation space with the goal of creating a long-term business. For example, \Pthree{} had attempted several previous start-up ideas, saw the funding success of CivitAI, \pquote{they had raised fifteen million dollars,} 
and pursued a new venture 
\pquote{that really lit the entrepreneur in me to say I need to jump on this.}} 
 Both participants (\Pone{}, \Pseven{}) who had experience with online sex work were interested in using AI to support content creation
\pquote{so I don’t have to pay someone to take photos of me in my underwear}~(\Pone{}) and to \pquote{fill some social media gaps}~(\Pseven{}). However, the tools they used were not able to accurately and consistently depict them visually. While most did not mention integration with the existing sex industry,
entrepreneurs \Ptwo{} and \Ptwentynine{} were interested in partnering with sex workers to help them generate content since
\pquote{one of the pain points for them is creative burnout}
~(\Ptwo{}). \Ptwentynine{}~recalled receiving a commission request from a sex worker, \pquote{I guess she was a pregnant woman... 
And she wanted me to create, you know, AI content
probably because she... couldn’t maintain that body type because she was probably going to give birth.}

\subsection{\new{Socialization}}
\new{The desire for community also motivated creators to both join AIG-SC spaces and be active contributors among their peers.}
For many participants, AIG-SC communities enabled them to build new friendships, discuss sexual interests, or find purpose through community leadership. As \Pthirtyone{} summarized: 
\begin{quote}\pquote{[People] share their images and kind of make friends with each other, share tips like recommendations to other models or maybe other programs to run those models. And so it’s like a... community of the art that they make and how to make it.}
\end{quote}

AIG-SC communities also enabled a unique space for people to share sexual content without stigmatization:
\pquote{I just like showing [AIG-SC] to strangers who are there for that stuff. Because otherwise... my friend’s like,
`Oh, what have you been doing?'
`I’ve been making pornography.' Like, that’s a bit weird [laughs]}~(\Ptwentysix{}). \new{These spaces can act as} affinity groups for people with diverse sexual and gender presentations. What drew \Pthirtyfour{} into their online community was that it was:
\pquote{basically a bunch of gay nerds that were like, `Well, none of these tools for making big boobs are working for making dicks. So, let’s see what we can do'}~(\Pthirtyfour{}).
However, despite this, several participants still ended conversations with the research team by remarking on the interview as a rare opportunity to speak openly with someone about this and the enjoyment it brought:
~\pquote{I don’t share this with anyone close to my personal life. This is... a weird hobby/gig thing I’m kind of moonlighting with. So, it’s always fascinating to have an opportunity to kind of discuss it}~(\Ptwentynine{}).

\subsection{Anti-Censorship and AI Safety}
\label{sec:ai-censorship-and-safety}
Some participants mentioned that they created AIG-SC \revision{as a form of political activism (e.g., hacktivism~\cite{jordan2004hacktivism}) against the content restrictions enforced by AI companies}. Others engaged in similar behavior but with a different goal: to improve AI safety through company-coordinated red-teaming.

Several participants learned how to make AIG-SC via jailbreaking and/or developed local systems to sidestep perceived censorship and restrictions.
Like several others,
\Pthirtytwo{} held general disdain for AI companies and their \pquote{hard censorship... to avoid NSFW content}; because of this, they created and shared jailbreaking guides to ensure \pquote{that everyone can have access to every tool they need to enjoy an uncensored experience with AI} and to defy corporate censorship:~\pquote{What I enjoyed the most? Sharing knowledge. 
...And also helping break big AI company TOS [Terms of Service].}
For some, this social prestige of standing against companies was a strong motivator. \Pthirtynine{}, a self-described prominent jailbreaker, noted that they enjoy jailbreaking \pquote{whatever [system] people are having the trouble with the most... another major Claude jailbreaker that kind of went on break, so they [needed help]. It's like, all right, all right, all right. Let me take care of this.} 
Beyond helping others in the community, \Pthirtynine{} said that they enjoy \pquote{being regarded as a badass--I like knowing everything and I like people knowing that I know everything.}
Others got around censorship by learning how to produce content locally, likening themselves to digital pirates:~\pquote{I’m not going to pay a lot for that thing. I’m going to figure out how to do it myself. So there’s a sort of pirate aspect of it}~(\Pthirtyfour{}).
\new{Local generation was also used to circumvent potential copyright violations:
\pquote{You can’t generate a Marvel character... why we are choosing to run our own models is we can actually generate what we want to generate. We’re not getting blocked by these big companies} (\Ptwelve).}

Other participants purported to generate AIG-SC and jailbreaks to understand system boundaries in order to better secure AI systems.
For instance, \Pfortyfive{} worked with an AI security provider to red team their systems and report what is possible.
\Pfortyfive{} claimed that their jailbreaks were reported back to the vulnerable companies, allowing them to improve their models' safety \pquote{because I am also worried about the bad impact of artificial intelligence out of control... the current development of artificial intelligence is like an arms race. For speed, safety studies may be lagging.} 
\Pfortyfive{} explained that 
\pquote{porn is the most common [censored topic], so I chose this topic [often]} for jailbreaking, but they also mentioned generating child harm and suicide content because \pquote{if you can make the model generate more harmful content, it means that you don't need to test simple topics.} \revision{However, we note that the same participant also disclosed creating NCII (Section~\ref{sec:AIG-NCC}), highlighting the need for responsible red-teaming guidance (Section~\ref{sec:discussion}).}

\Pfortytwo{} noted that the broader AIG-SC space includes many folks like \Pfortyfive{} who work with companies:~\pquote{A bunch of them are actually red teamers. Like a shocking amount. Like [one member]...he's been inactive because he's doing Claude Red teaming now.}
We found a spectrum of beliefs regarding how effective jailbreaks should be handled.
While \Pfortyfive{} kept jailbreaks between themselves and the vulnerable company, other red teamers do not. \Pfortytwo{} noted that the Claude red teamer still provides jailbreaks for others:~\pquote{every once in a while, he drops in and drops us something that we can use just for fun.} 
While a believer in uncensored AI, \Pfortytwo{} noted that through their jailbreaking practices over the years, they have likely, if inadvertently, improved the system's security: \pquote{I'm certain that... just our existence on the Claude website and other websites has probably advanced security measures a significant amount.}

\section{AIG-NCC Perpetration and AIG-SC (\ref{rq:AIG-NCC})}
\label{sec:AIG-NCC}
\label{subsection:NCII}
\new{AIG-NCII is a form of image-based sexual abuse~\cite{mcglynnImageBasedSexualAbuse2017} and can cause severe mental health consequences, reputational damage, and financial harm~\cite{mcglynnItsTortureSoul2021, batesRevengePornMental2017, brighamViolationMyBody2024, flynnSexualizedDeepfakeAbuse2025}.}
\new{AIG-NCII can also occur as an extension of intimate partner violence as a form of control and humiliation~\cite{flynnSexualizedDeepfakeAbuse2025}.}
\new{Public figures, particularly women, that are not personally known to perpetrators are frequently targets, such as influencers~\cite{kattenbargeDeepfakePornTikTok2023}, celebrities~\cite{nadeembadshahNearly4000Celebrities2024, gracepanettaTheySoldMy2026, maibergAIGeneratedTaylorSwift2024}, and politicians~\cite{barbararodriguezAIEntersCongress2024}.}
\new{Indeed, in line with prior work~\cite{flynnSexualizedDeepfakeAbuse2025},  participants disclosed or observed others creating AIG-NCII that depicts known individuals (e.g., friends, family members, peers) and public figures.} 

\new{This section further documents the motivations and creation methods of AIG-NCII perpetrators who participate in the same online spaces as non-perpetrators. This section is informed by accounts from three participants who disclosed AIG-NCII perpetration, as well as narratives from others who either received AIG-NCII requests or encountered perpetrators.}

\new{We emphasize that many participants stated personal boundaries against creating AIG-NCC. For example, \Pfour{} suggested the need for greater accountability:  \pquote{you have to take responsibility... you're not going to go to jail for doing upside down lesbian elves but you will if they’re under eighteen.}}
However, three (\Peleven{}, \Pfortyfive{}, \Pfortysix{})  disclosed that they created NCII,\footnote{We define NCII based on U.S. federal law~\cite{tedcruzS146TAKEIT2025} but note that participants who reside elsewhere are subject to different local laws.} another (\Ptwelve{}) disclosed that their friend created NCII, and two (\Ptwo, \Ptwentynine{}) received commission requests for AIG-NCII that they denied. None of our participants disclosed creating AIG-CSAM.
We anticipate under-reporting due to social desirability bias and concerns over possible legal ramifications.

\subsection{What Motivated Perpetration and How Was Perpetration Rationalized?}

\new{Given our sample size, the motivations listed are not exhaustive, and while} we encountered varying self-described motivations for creating NCII, which further indicates the need for a broad range of interventions. None of the participants described themselves as being motivated to cause interpersonal harm \new{(though we again note that this may be due to desirability bias)}. Therefore, 
participants who created NCII \new{downplayed their actions and} did not comment on the harms of NCII or signify belief that such behaviors are harmful.\footnote{\Pfortyfive{} alluded to NCII as being ``unsafe'' but the focus of their comments was on their belief that ``\textit{pornographic}'' content is ``unsafe.'' They did not specifically articulate any harms that related to NCII.} In one case, \Pfortysix{} later tried to deny that they had admitted to creating NCII, indicating \textit{some} concern about its moral acceptability. \new{Overall, we observed various forms of rationalization from perpetrators and some non-perpetrators. As other scholars have suggested, this may be due to the accessibility of content creation and a perceived lack of consequences~\cite{flynnSexualizedDeepfakeAbuse2025}. Such rationalizations} further underscore the need for clearer norm-setting (Section~\ref{sec:discussion}) through community-based interventions, more education around the harms of NCII, and legal regulations addressing AIG-NCII.

\header{Sexual Gratification, Novelty, and Community}
\Ptwelve{} described that their friend created NCII for sexual gratification: \pquote{he knows in real life that he’s not able to get with maybe a crush or something like that. The next best thing you can do is generate an image.} Therefore, the friend created \pquote{insanely realistic results of his friends in person. But it like looks just like them, but they’re anime characters.}\footnote{While not all photo transformations retain another's likeness, we consider this AIG-NCII as the target's likeness was explicitly retained.}.
While they noted that more photorealistic content \pquote{could be a little creepy,} they later described this as \pquote{gray areas, where people are like, `Shit, am I supposed to be able to generate this?'} They then reasoned that people \pquote{generate locally} to evade scrutiny and that 
what people view privately does not affect others, a belief held by nearly 50\% of U.S. residents surveyed in prior work~\cite{brighamViolationMyBody2024}.

Outside of sexual gratification and consistent with Flynn et al.'s findings~\cite{flynnSexualizedDeepfakeAbuse2025}, which were based on 10 interviews with AIG-NCII perpetrators, curiosity and peer bonding motivated perpetration. For example, \Pfortysix{} disclosed that they \pquote{got bored with porn} and wanted to \pquote{try new things.} In addition to corporate red-teaming, the jailbreaker also mentioned: \pquote{I sometimes get curious. Build a porn scene for two celebrities}~(\Pfortyfive{}). Such ``curiosity'' is then facilitated by the ease of AIG-NCII creation~\cite{flynnSexualizedDeepfakeAbuse2025}.

Creating and sharing NCII can be a form of bonding that emerges out of a desire to perform for, be accepted by, or gain status in a group~\cite{flynnSexualizedDeepfakeAbuse2025, hanCharacterizingMrDeepFakesSexual2025}.
One participant (\Pfortysix) characterized themselves as contributing to an online community by creating and sharing a LoRA depicting a new celebrity that others previously had not thought to create, much in the same spirit of Mr. Deepfakes~\cite{hanCharacterizingMrDeepFakesSexual2025,timmermanStudyingOnlineDeepfake2023}. They disclosed that, \pquote{I was just experimenting. There was this celebrity. Her LoRA [didn't] exist on [the] web. So, I thought I should create it on my own.} They posted the LoRA on a public forum (which has since banned NSFW content). As \Pfortytwo{} described, \pquote{There's one guy in the [community] who just keeps making stuff of [famous actress]... all he does is just make [nude] pictures of her.}

\header{Financial Opportunities Amid Lack of Regulation}
Monetary incentives also exist for creating NCII. Several participants reported either creating or receiving requests to create AIG-NCII of someone they did not know, but who was another real, identifiable person.
\Ptwentynine{} mentioned rejecting paid requests, and that one requestor even asked for content depicting \pquote{a relative, like, the guy’s mom or... maybe his sister? And he is like, `Can you do sexual content based on this person?' And I’m like, `Absolutely not.'} 
\Ptwentynine{} hypothesized that they received the requests due to their hyperrealistic content style, but added, \pquote{I’m sure that other content creators probably get a few requests.}

Those who do not share the same boundaries are readily taking NCII commissions~\cite{hanCharacterizingMrDeepFakesSexual2025, dawoudUndergroundMainstreamMarketplaces2026}. 
\Peleven{} was paid to help a client create a LoRA that depicts a specific person. Their client gave images of a specific person unknown to them: \pquote{I don’t know where [my client] found this face, but he just gave me a folder of this person’s face and I was like, `Okay, I guess we’re going to use this person,' so we took their likeness and... trained a LoRA on them.}
When asked for more detail, they treated the instance purely as a business transaction, \new{without acknowledging that this constitutes NCII} and \pquote{never followed up} on what happened with that LoRA after they provided it.

\Ptwo{}, a developer who \new{rejected AIG-NCII requests} and prided themselves on building an AIG-SC system with principles of \pquote{ethical consumption,} noted that the lack of regulations allows companies to profit from enabling NCII:
\pquote{[There's a] lack of ethics because there's no regulatory constraints... from a company’s standpoint, that’s a quick cash grab.}
% They’re not going to exist long-term.}   
For their own text-to-image AIG-SC generation tool, they noted that: \pquote{The number one requested feature... is an undressing feature where you upload a photo of anybody, and then it undresses them. People have emailed me saying, `Hey, are you going to do this?' And I say, `Absolutely not.'}

\header{Boundary Pushing Without Boundaries}
Lastly, \Pfortyfive{}, a professional red teamer, claimed that they created and posted NCII on social media\footnote{In a few instances, participants shared their public profiles and posted content with us. In this case, we found it contained NCII.
In line with our consent form and the community rules from which we recruited that ban the posting of NCII, we reported the publicly posted NCII to the platform. See Ethics in Appendix~\ref{sec:ethics} for details.}
because they \pquote{want people to know that the model is not as safe as imagined.} However, their posts typically contained no or minimal information (e.g., just the name of the model). Without additional messaging in line with their stated intention, the posts are decontextualized pieces of NCII for those who stumble across it, which contradicts their statements on exposing the \pquote{bad impact of artificial intelligence out of control.}

\subsection{What is Made and With What Systems?}
\new{Prior work~\cite{gibsonAnalyzingAINudification2025, flynnSexualizedDeepfakeAbuse2025, williams_there_2025}, media coverage~\cite{emanuelmaibergInstagramsNudifyAds2024, murphyNudifyAppsThat2023, katecongerElonMusksGrok2026, lola_murti_tech_2026}, and regulatory efforts~\cite{pieterhaeckEUSetBan2026, jasmine_mithani_minnesota_2026} have focused on pre-built ``nudification'' or ``undressing'' apps and services that are available at low (or no) cost and require no specialized skills.
While one participant (\Pfortysix) had used a nudification app, more skilled creators, such as our participants, utilize open-source models and jailbreaks in commercial models (e.g., DALL-E, Midjourney) to create AIG-NCC.}
We describe direct modes of creation disclosed to us below. In addition, one participant (\Pfortysix) alluded to interacting with Telegram bots~\cite{burgessMillionsPeopleAre2024} that aid NCII creation, and, as mentioned, two others (\Ptwo, \Ptwentynine) received paid NCII requests from others (which they declined).

\smallskip
\noindent
\textbf{Face-swapping algorithms}, where existing faces are transplanted onto existing nude images, were used by \Peleven{} who noted that \pquote{the face swap technology is really good.} As explained by \Pthree{} (who had not created NCII), \pquote{so this is the original video, right?... It’s the exact same video... but now it’s got this different face.}

\smallskip 
\noindent 
\textbf{Jailbreaking.}
Since text-to-image models (e.g., DALL-E, Midjourney, Stable Diffusion) learn information about public figures during the training process, they can be jailbroken to create NCII, as \Pfortyfive{} had done to create NCII depicting a famous political figure and celebrity.

\smallskip 
\noindent 
\textbf{Fine-tuned machine learning models} were used by \Pfortysix{} to fully generate new images in a person's likeness.
Fine-tuning is often implemented by training a Low-Rank Adaptation (LoRA) on existing models to generate a specific fictional, or real, identity.

As described by \Pthree{}, \pquote{LoRAs are the mods... you can mix and match. You can say here’s a character and they’re doing this;} perpetrators used the same technique to generate NCII of a real person. To perpetrate NCII, \Pfortysix{} mentioned that \pquote{You just need 25 to 50 images of a celebrity or a person [for the] LoRA you are trying to make... it's very easy. You just click on run, and there you go. You [then] enter your prompt and it will generate [an] image.} 
\Ptwelve{}, who mentioned that their friend created anime-style AIG-NCII, disclosed that the friend was able to do so by training a LoRA on images of his peers.

As machine learning techniques improve, fewer images are needed to train a LoRA to depict a specific person. Through the course of this research, we encountered methods advertising the ability to create a LoRA using a single image by leveraging one-shot learning~\cite{life-feiBayesianApproachUnsupervised2003}. Researchers developing one-shot learning techniques should consider how their work can further lower the barriers to NCII creation, particularly since perpetrators have been shown to leverage findings and public codebases from academic papers~\cite{hanCharacterizingMrDeepFakesSexual2025}.

Participants also mentioned that combining techniques can account for individual techniques' limitations: \pquote{you could train a LoRA which only gets you so far---it generates accurate hair, but the face... you pick up on such little details if something is different. So then you’d also use the LoRA plus a face swap
...[that's] a pretty powerful technique}~(\Peleven{}). Once a LoRA is created, it can be shared openly on platforms and allow others to create limitless NCII of a specific individual.
Until a recent ban~\cite{maibergCivitaiBanReal2025}, CivitAI was commonly used to share LoRAs that nonconsensually depict celebrities, influencers, and others~\cite{emanuelmaibergAIPornMarketplace2023}.

%% file: arXiv-sections/discussion.tex
\section{Discussion}
\label{sec:discussion}

\subsection{Reflections on the Impacts of AIG-SC}
\label{subsec:impacts}

Through our conversations to understand who, what, and why people generated AIG-SC (\ref{rq:background}, \ref{rq:content}, \& \ref{rq:motivations}), we find that AIG-SC has several self-proclaimed beneficial use cases for creators and consumers (Section~\ref{sec:usecases}). In addition to appealing to sexual interests, AIG-SC was described as allowing sexual and gender self-discovery, creative artistic expression, community building, and technical learning. 
Yet there are also societal implications (outside of AIG-NCC), including labor displacement, content theft, and the perpetuation of unhealthy norms, which some participants offered reflections on.

\header{Sexual Content Norms} 
\new{Participants noted that AIG-SC content is predominantly shared online by \pquote{cisgender straight men, which reflects on `male gaze'-y types of thing}~(\Pthirtysix). This leads to content that frequently depicts women with unrealistic body proportions, which \Pseven{} characterized as \pquote{Barbies gone wrong... I know very few girls that are shaped that way,} and \Pten{} described as a \pquote{misogyny focus on the female body.} They went on to say that \pquote{most of the generated images online are kind of the same. You make a very big breasted
woman, naked and in the same position as the other seventy ones.} } 
\new{While creators can use AI to enable new forms of content creation, other scholars have similarly suggested that there is the potential for AIG-SC to reinforce harmful gendered stereotypes, normalize highly exaggerated body proportions, and lead to content that is increasingly dehumanizing~\cite{depountiIdealTechnologiesIdeal2023, lapointePresentFutureAdult2025}}.

AIG-SC allows for content to be created in situations or physical settings that are, as writer Leo Herrera found, ``impossible for average porn creators to capture: stadiums, airports, police stations, graduations, weddings and marathons''~\cite{herreraSyntheticGenres22025}. Since new content can be generated without physical human involvement, scenes that push the boundaries of humiliation or violence, in scenarios that few would consent to, can now be made using AI.
 Recent calls in psychology~\cite{grubbsPornographyUsePsychological2021} and public health~\cite{grubbsPornographyUsePublic2023} lament a lack of mainstream consideration of the effects of sexual content in their fields. We amplify their call for scientific inquiry on the negative and positive effects of AIG-SC, in particular, on individuals, their relationships, and society. 
Such inquiry is necessary to find a balance between AIG-SC's virtues and harms.

\header{\new{The Sex Industry}}
AIG-SC and its increasing monetization have direct implications for the sex industry. 
As AI continues to advance, it could pose financial threats to sex workers if AIG-SC focuses more on replacing human-made content, rather than supporting its creation.
As discussed in Section~\ref{sec:motivations}, AIG-SC may also shift demand for content as some participants who desire more niche content replace their content consumption with AIG-SC.
Increasingly, AI-generated content may emerge on the same platforms as human-made content (e.g., AI-generated accounts on OnlyFans~\cite{emanuelmaibergPeopleAreUsing2025, emanuelmaibergBoomingAIPimping2024, dawoudUndergroundMainstreamMarketplaces2026}), as described by \Peleven{}, who trained a LoRA to produce content for OnlyFans such that viewers believe, \pquote{'This is a real person.' [chuckles] That was the whole goal}~(\Peleven). As previously mentioned in Section~\ref{sec:AIG-NCC}, this LoRA was trained using someone else's likeness and constitutes NCII.
Outside of OnlyFans, \pquote{the emerging thing is AI influencers, AI models but [on platforms] friendly to AI content}~(\Pthree).

Although AIG-SC and its increasing monetization hold implications for the sex industry, many participants felt that the content they created is orthogonal. As \Ptwentynine{} exemplifies, 
\pquote{I’m not trying to replace, like, real life pornography... [this is for] the novelty of it ...the images I’m posting, they’re all different, randomly generated unique people... what I’m doing is very separate from, like, mainstream porn industry.}

Only one participant (\Ptwo{}) mentioned stolen labor as a concern in the development of AIG-SC. Meanwhile, \Pseven{}, a sex worker who produced digital content, characterized those who are monetizing AIG-SC as outsiders who,
\pquote{don’t understand the sex worker business... not realizing the work that this job really entails}~(\Pseven). They therefore called for fair compensation and authentic partnership on tools built collaboratively with sex workers, adding that, 
\pquote{...if you want to use our images to be trained on AI, we have to be paid for that... there’s scum-sucking companies that want to try to steal our images... if we partner with AI companies that are truly trying to benefit us... we can make money off of it, as long as it protects us.}

As \Pseven{} noted, AIG-SC, whether monetized or not, is often trained using data that was obtained nonconsensually, including the use of content produced by sex workers without appropriate compensation. AIG-SC is also bringing in individuals with no experience working in the sex industry, which \Ptwo{} characterized as: 
\begin{quote}
\pquote{people who can slap together a website... probably with the expectation of, `This is going to fail within a year.'... I think the[se] software engineers do know what they’re doing is not ethical, and they’re like, `I’m going to make money off of it anyways.' I think the majority of [AIG-SC entrepreneurs] don’t really understand the implications of what they’re doing.}  
\end{quote}

As technology advances even further, there will likely be increasing concerns around labor displacement as well as economic pressures for sex workers to integrate AI into their work. 
For example, \Pseven{} shared concerns that companies are trying to retain their likeness and generate images against their interests, violating their rights and financially impacting them:
\pquote{It got snuck in a lot of our contracts performing with mainstream companies... they were trying to have us sign our AI rights to that company, for that video. And that directly affects [us]---if a company can use my likeness and not pay me.}
 We therefore echo \Pseven{} and others' calls~\cite{butlerSexWorkerHuman2025} for direct partnerships with sex workers and for future work that centers sex workers' experiences.

Lastly, we emphasize that the technical capabilities that power AIG-SC were created by exploiting sexual content that was scraped from the internet. This training process has no regard for consent~\cite{princessacintaqiaStopNonconsensualUse2025}, with prior work documenting the inclusion of CSAM and NCII~\cite{thielGenerativeMLCSAM2023} in training datasets, and of course the appropriation of sex workers' labor without appropriate compensation. We echo calls for future work developing systematic approaches to ethical collection of training data with compensation and consent~\cite{princessacintaqiaStopNonconsensualUse2025}, building off existing efforts ethically assembling evaluation datasets~\cite{xiangFairHumancentricImage2025,scheuermanResponsiblyTrainingFoundation2025}.

\header{\new{Creative Work}} Participants held disparate beliefs on the impact of AIG-SC on artists, writers, and human creativity. Some creators came from creative backgrounds (Section~\ref{sec:creator-backgrounds}) or described themselves as creating art (Section~\ref{sec:usecases:artistic}) but noted that, \pquote{there's this battle between artists and AI users}~(\Pten). Accordingly, a few participants noticed that in online spaces that heavily include artists or artistic practices (e.g., communities that create and share sexual art and writing), AI content is not welcome: \pquote{Everyone's seen a whole bunch of AI-generated images. Some people are sick of them} (\Psixteen). 
\Ptwo{} initially posted their AIG-SC content to hundreds of forums that allow sexual content sharing but experienced pushback, \pquote{[some] were like, `This is AI, [we] don’t allow this shit.' So, people get angry online... I think the realness factor is important for some people.} 
Meanwhile \Pthirtyseven{} found that \pquote{the furry community is really negative about AI... certain sites even have a ban against AI.} Given that generative AI enables large volumes of sexual content to be easily created and shared, further work should explore how the encroachment of AIG-SC content affects communities where it is unwanted.

Although participants were aware of tensions around AI-generated content and that \pquote{there's like a division here, that  people who do handmade art, like drawn on tablets and stuff, kind of feel betrayed by AI} (\Ptwentysix), they mostly held positive sentiments around AI (as users of it themselves) with the belief that it is \pquote{just another tool and you can collaborate with it} (\Ptwentysix).
For \Pnineteen, who creates AIG-SC text content, this led to personal conflict as they commented on the potential impacts of monetary loss for writers,
\pquote{so I'm not taking the market away with actual writer. But I am sometimes scared that artist would lose their job/passion if AI becomes too good... I guess I live with the hypocrisy for now}.
On the other hand, \Pnine{} resented being characterized as \pquote{`a bad person'---that’s what I dislike the most} because they use generative AI and argued that, \pquote{it's way cheaper to use a gen AI model than to hire an artist. And of course it’s way faster too. And that's where we come in handy.} Similarly, \Pten{} commented that, \pquote{I believe if I can make it easier, it's better} and rejected the notion that they are \pquote{stealing [artists'] work... the technology itself learns the same way as the person do, only faster, it’s just natural for it to exist}. \Pthirtyseven{} echoed this sentiment and argued that AI artists and traditional artists both \pquote{enjoy the process of creating something from pretty much nothing}.
However, AIG-SC creators sometimes fine-tune models directly using the styles of existing artists, which \Pten{} commented on. They believe that, \pquote{it’s very unethical to start selling people things with the signature of a famous artist... sometimes I try not to use directly the style of artists I know or I appreciate. Or just mix them.} Similar to sex workers' concerns, artists also fear labor displacement, the nonconsensual use of their content, and artistic forgery~\cite{kyi2025governance, jiang2023ai, goetzeAIArtTheft2024}.

\subsection{Addressing AIG-SC as Computer Scientists}
\smallskip\noindent\textbf{\facct{Does Model Openness Lead to Abuse?}}
\facct{In recent years, companies have begun publicly and irrevocably releasing open-weight models, with the stated goal not only of democratizing AI by allowing others to reuse and extend costly-to-train models~\cite{widder2024open} but also of enabling transparency, allowing for third-party assessments of system properties like fairness and security that regulators and academics have advocated for in other settings~\cite{costanza2022audits}. Others critique this openness, pointing out the significant threats to safety from harmful reuse and extension use cases such as the AIG-NCC use cases we documented~\cite{kapoor2024position}. Prior work attempting to adjudicate between these two stances has pointed out the lack of data on ``marginal risk''~\cite{kapoor2024position} -- rigorous documentation of realized societal harms and benefits -- that can inform decision-making about model openness in the future. Our results take a step toward filling this gap. } 
\facct{In particular, we find that open-weight models with LoRAs are central to both AIG-SC and -NCC creation, and we offer empirical evidence of the low technical barrier to reuse and extension of open-weight models: AIG-SC communities support rapid upskilling among members, with some holding PhDs and having published peer-reviewed machine learning research. We observe few barriers (technical or otherwise) preventing the generation of AIG-SC, or -NCC.  
Given the lack of technical ability to prevent AIG-NCC while allowing for AIG-SC or AI generation of any images of humans in general~\cite{chandraReducingRisksPosed2024,kapoor2024position}, it remains to be seen whether benefits from the, now-irrevocable given the release of open-weight models, decision to deploy technologies that generate images of humans will outweigh the detriments.}

\facct{Furthermore, we also find that red-teaming communities may significantly overlap with AIG-SC communities, and methods developed in corporate-sponsored red teaming can then be shared and actively exploited among a broader community that contains abusive actors.
These findings underscore the urgent need to answer questions about unclear norms for external AI red-teaming~\cite{longpreSafeHarborAi2024} including questions industry has left unanswered such as ``Who should red team and why?'' and ``What protections should we put in place to ensure the safety of the red team?''~\cite{ganguliRedTeamingLanguage2022}. We must also understand how to craft safe harbors that protect good actors~\cite{longpreSafeHarborAi2024,pfefferkornTheresOneEasy2026} but not harmful ones operating under the auspices of security. 
Such inquiry should revisit existing questions about the efficacy and ethics of incentivizing vulnerability-finding by freelancers~\cite{egelmanMarketsZerodayExploits2013}, and further investigate~\cite{sinhaWhatCanGenerative2025} whether and how suggested best practices from cybersecurity programs~\cite{piaoUnfairnessBugBounty2025,votipkaHackersVsTesters2018} have been implemented for AI red-teaming, to identify gaps in the efficacy of these strategies for guiding ethical behavior by AIG-SC-focused red teamers. 
New methods focusing on improving AI safety should critically examine who benefits from existing AI transparency and accountability practices.}

\smallskip
\noindent
\textbf{Parallels between AIG-SC and Hacker Cultures.} 
We find parallels between AIG-SC creators and prior literature on hackers that can inform future work. Like hackers, the AIG-SC communities we study exhibited elements of being a subculture including ``a
social scene, slang, and value system that defines boundaries''~\cite{holtHacksCracksCrime2005} between them and general internet users. The frequency with which many participants created AIG-SC echoes hackers' ```easy, if not all-consuming, relationship'...with technology''~\cite{holtHacksCracksCrime2005} and our participants exhibit a similar set of values to those expressed in the Hacker Ethic~\cite{levyHackersHeroesComputer1984}: that computers create art and beauty and can change your life for the better, as well as anti-censorship beliefs and a mistrust of authority~\cite{holtHacksCracksCrime2005}. Further parallels with prior work on hackers can be seen in our participants' strong degree of curiosity about technology (in our case, AI), desire to build social connections, and engagement in status-building through demonstration of mastery (by sharing high-quality, or, concerningly, extreme AIG-SC)~\cite{holtHacksCracksCrime2005,xuWhyComputerTalents2013}. There exists a spectrum of ethics in hacking behavior (from white to black hat), with prior work finding the primary delineator is an individual's ``moral values and
judgment''~\cite{xuWhyComputerTalents2013}. 
Computing research has focused primarily on white-hat hackers~\cite{walsheLongitudinalStudyHacker2022,piaoUnfairnessBugBounty2025,votipkaHackersVsTesters2018}. We amplify existing calls for research into a wider range of adversarial behavior, with a focus on initiation~\cite{xuWhyComputerTalents2013}. 
Such inquiry should extend prior work on general attitudes toward AIG-NCII~\cite{brighamViolationMyBody2024} and mediators of IBSA perpetration~\cite{henryImagebasedSexualAbuse2024,flynnSexualizedDeepfakeAbuse2025} to investigate the attitudes, social influences, and experiences that mediate (un)ethical AIG-SC creation.

\smallskip\noindent\revision{\textbf{Community-based Interventions.}} 
Interventions that work on less savvy perpetrators who rely solely on nudification tools~\cite{flynnSexualizedDeepfakeAbuse2025}, such as deplatforming those tools, will be less effective in preventing AIG-NCC creation among those with our participants' skills. 
\facct{We offer an individual-focused analysis of members of AIG-SC communities as a first step to taking an institutional lens~\cite{ostromBackgroundInstitutionalAnalysis2011} to this ecosystem. This focus allows us to map the individual-level ``Action Arena'' (actors and their actions) of AIG-SC creation within these communities, detailed in Appendix~\ref{appendix:IAD}. 
Future work should focus on extending this understanding to community-level analysis of norms and governance. Such work is critical given that} \new{prior work has found social influence to be a significant mediator of ethical vs. unethical hacking and IBSA perpetration~\cite{xuWhyComputerTalents2013,henryImagebasedSexualAbuse2024}.}
While \new{many} participants expressed boundaries against creating and sharing AIG-NCII, a few rationalized abusive content and engaged in its creation (Section~\ref{sec:AIG-NCC}). Further work is needed to understand the challenges of governing AIG-SC online spaces so that interventions can be crafted to prevent the influence of AIG-NCC creators and their content on AIG-SC creators. Such work should also examine how perpetrators discover AIG-SC spaces and how creation methods overlap between abusive and non-abusive use cases.
Further, given that three participants had completed CS or AI coursework or degrees and eight were employed in the tech industry, computer science educators and software engineering professionals should also intervene, e.g., via CS ethics interventions~\cite{kohnoEthicalFrameworksComputer2023} focused on AIG-SC and -NCC.

We note that all such work must be done with care and a focus on protecting, not censoring, free speech, while simultaneously advocating strongly against abusive behavior. General censorship of non-abusive sexual content communities and generation can push non-abusive creators into even more abusive spaces (e.g., 4chan), increasing rather than decreasing exposure to harmful social influences. As one AIG-SC community moderator explained,
\begin{quote}
\pquote{We had news articles come out... lumping us in with [AIG-NCC]... which is very much not what we stand for... there is a nuance to [AIG-SC], not everyone is using it for this nefarious purpose}~(\Pthirtyone{}). 
\end{quote}

\smallskip
\noindent\textbf{Addressing AIG-NCII in a Global Context.} 
% - Trying to stop AIG-NCII by removing capabilities to generate sexual content is not going to stop AIG-NCII. need to be thinking about AIG-NCII through the lens of consent, rather than just sexual content (AIG-NCC is broader than sexual content)
% - content is often nonconsensually used in the creation of this technology, consensual sexual content is very context-dependent.
\facct{While this work discusses the creation of \textit{sexual content} and highlights harms as a result of the nonconsensual creation of \textit{intimate content} using AI, sexual content is only a subset of intimate content. Definitions of NCII are highly contextual and cultural---in certain cultural contexts, NCII can include clothed visual content in romantic settings~\cite{batoolExpandingConceptsNonConsensual2024}. In order to holistically combat abuse, solutions for mitigating AIG-NCII must fundamentally center on \textit{consent} and the nonconsensual use of individuals' likeness.}
% rather  than solely focusing on nude and/or sexual content.}
% \smallskip
% \noindent\textbf{\revisionthree{Moderating AIG-SC Globally.}} 
% \revisionthree{
\facct{Laws must also catch up to the global context in which AIG-NCII is created.
As evidenced by our study, 
AIG-SC creators are receiving commission requests for NCII, even when they do not advertise AIG-NCII creation. While our participants declined these requests, regulations must reflect the global marketplace for AIG-NCII creation, which enables perpetrators to commission AIG-NCII from individuals who reside in a different country, whether unintentionally or purposefully to evade laws. 
However, to the best of our knowledge, the United Kingdom is the first country to criminalize \textit{commissions} for AIG-NCII, even if creation occurs outside of the country~\cite{huiWhatKnowUK2026, parliamentoftheunitedkingdomDataUseAccess2025}.}

\smallskip
\noindent\textbf{\revision{Sex-Positive Computer Science?}}
To engage effectively in setting positive norms, research and even red-teaming~\cite{gillespieAIRedteamingSociotechnical2024}, computer scientists must gain comfort in addressing sexual content and related topics. As noted in Section~\ref{sec:relatedwork},
sex-positive approaches recognize that sexuality is a natural human experience that is deserving of respect, acceptance, and discussion (note: \textit{it is not} the unconditional support of sexual activities). Due to improved outcomes, sex-positive approaches have increasingly been incorporated in sexuality-intersecting professions, including general healthcare~\cite{nimbi2022biopsychosocial},
 social work~\cite{williamsResolvingSocialProblems2013}, clinical and counseling psychology~\cite{williamsResolvingSocialProblems2013, sprott2023clinical, herbitter2024mental, pillai-friedmanBecomingKinkawareaNecessity2015, alexander2019sex},
counseling for victim-survivors of sexual abuse~\cite{baggettSexpositiveAssessmentTreatment2017, ocallaghanThinkJustHaving2024},
and the creation of sex education and policies that are more effective in reducing perpetration, and perpetration-supporting beliefs~\cite{williamsResolvingSocialProblems2013, williams2015moving, peterson2024can,sower2023kink, carmody2005ethical}.
To combat this, a litany of guidelines and training programs have been created to help sexuality-relevant professionals become ``sex-positive'' and ``kink-aware'', and improve patients' outcomes~\cite{alexander2019sex, moors2025advancing, pillai-friedmanBecomingKinkawareaNecessity2015, sprott2023clinical,constantinides2019sex}. In the context of NCII, suppression of sexual expression undermines efforts to combat abuse by directly reinforcing societal norms that cause harm~\cite{dvoskinSpeakingBackSexual2023}. For example, the societal stigma of sexual content creates shame that leads to victim-blaming attitudes that minimize harm, excuse perpetration, and prevent victim-survivors from seeking help~\cite{flynnVictimblamingImagebasedSexual2023, eatonPerceptionsSexualizedDeepfake2026}.

In recent years, computer science and the subareas of AI, human-computer interaction, and computer security have increasingly intersected with uses and abuses relating to human sexuality. 
Based on this newfound intersection with human sexuality and the need to more holistically combat abuse, we ask: \textit{Is it time for computing professionals to become sex- and kink-aware? And what does a sex-positive approach to computing look like?}

%% file: appendix.tex
\input{appendices/community-architecture}
\input{appendices/institutional-analysis-and-development}
\input{appendices/methods-continued}

%% file: appendices/community-architecture.tex
\section{\facct{AIG-SC Community Ecosystem}}

\begin{figure}[h]
    \centering
    \includegraphics[width=\linewidth]{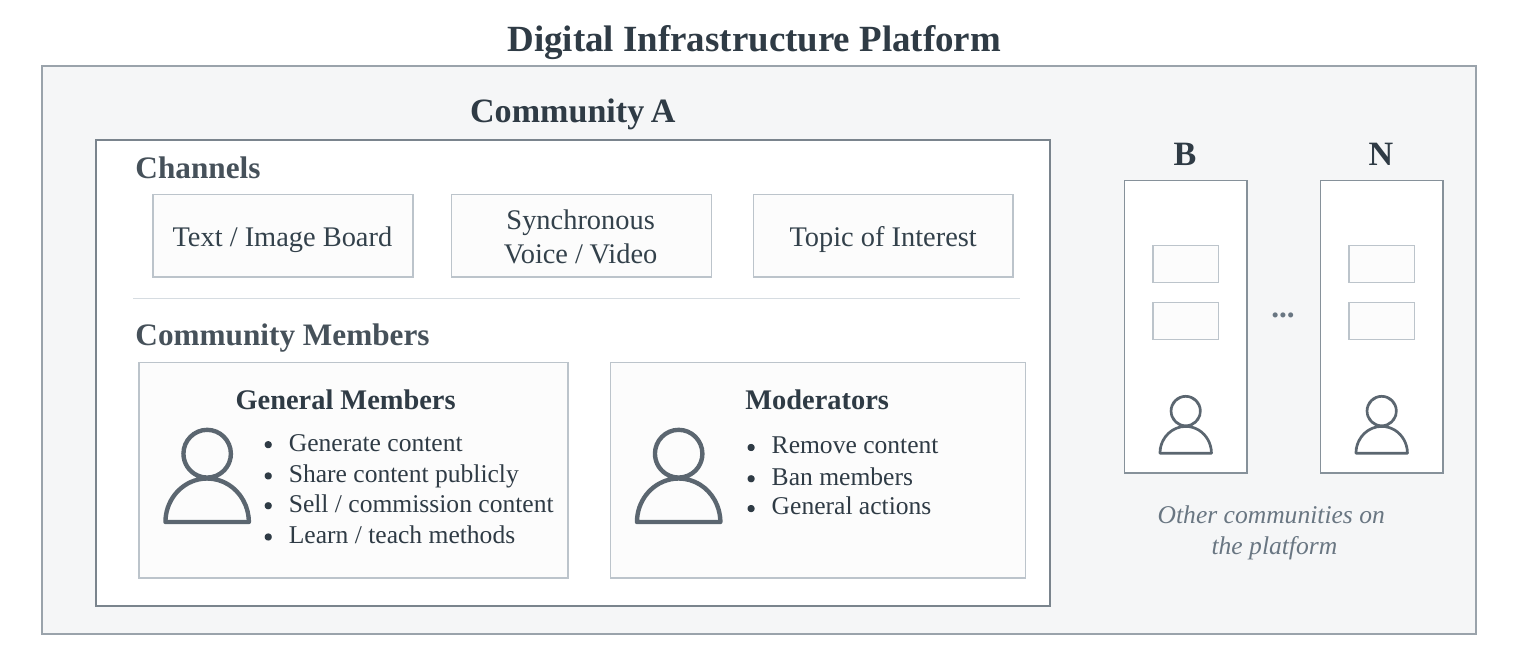}
    \caption{
        \facct{\textbf{AIG-SC Community Ecosystem} --
        \textmd{{\small 
      As noted in Section~\ref{sec:methods}, we find that AIG-SC content is often generated within a community-based ecosystem. Using existing digital infrastructure provided by a third-party ``Platform'', AIG-SC ``communities'' are often hosted within these platforms. Within these communities, individual ``Channels'' may be set up, often focusing on a particular topic or communication medium (e.g., asynchronous messages vs. synchronous audio).
        These communities are made up of both ``General Members'' who may generate content, post content, or commoditize their content among other behaviors. ``Moderators'' are privileged members who typically set and enforce governance actions (often content moderation) within the community.}}}
    }
    \label{fig:community-architecture}
\end{figure}

%% file: appendices/institutional-analysis-and-development.tex
\section{Institutional Analysis of AIG-SC Communities}
\label{appendix:IAD}
%explanation of IAD
% Utility of applying it to AIG-SC Communities
%Scope
The Institutional Analysis and Development  (IAD) Framework~\cite{ostromBackgroundInstitutionalAnalysis2011} is an analytical tool that can help explain emergent properties of interest based on the arrangement and makeup of major structures within institutions; these include: (1) exogenous variables that are imposed (e.g., environmental constraints, or rules), (2) the action arena, made up of the actors, their possible actions, what they value, and the interactions that occur between actors and actions, (3) outcomes of interest, and (4) a criteria to evaluate the institution upon (e.g., economic efficiency).
While traditionally applied to economics, it has also been practically applied to understand the effects of policies, such as the development of trustworthy AI systems~\cite{lewis2020rights}. Likewise, the approach can be used to understand how AIG-SC and -NCC content within communities may be affected by policies made by communities, platforms, and legal institutions. Here, we begin mapping our findings to the IAD framework, but emphasize that future work should extend this initial model to cover community-level analysis that can provide a more comprehensive understanding of this ecosystem.

\smallskip
\noindent
\textbf{Action Arena (Individual Actions).} 
%explanations of action arena
In Section~\ref{sec:usecases:sexualexpression}, we find a diverse set of actors who exist within AIG-SC communities with varied beliefs, incentives, and actions that they took. As such, we find that the action arena of AIG-SC communities is composed of several actors:

\textit{AIG-SC Members:} 
While most participants were members of AIG-SC communities, we find that these members' interests and motivations were often varied and included: personal sexual exploration, artistic creativity, financial opportunities, or anti-censorship ideologies.
These members could take actions such as privately generating content, sharing content publicly, selling or commissioning content, and learning or teaching about how to generate AIG-SC through pedagogy or distribution of resources.
Unfortunately, we also find that within AIG-SC communities, perpetrators willfully generate and/or distribute AIG-NCC and take much the same actions as general AIG-SC members.

\textit{Community Moderators:} Community moderators are select privileged members who, in addition to typical member actions and interests, can also take administrative actions such as removing content or banning members. Community moderators are significantly invested in the success of the community as a whole, either due to personal ideological beliefs or financial opportunities that may come from their involvement.
%different interests
% actions of each

\smallskip
\noindent
\textbf{Outcomes.} 
Outcomes of great concern within these communities, importantly, include the creation and sharing of AIG-NCC, including as a form of social bonding.
Other outcomes include learning -- the sharing of knowledge that enables AIG-SC or AIG-NCC generation -- and digital rights advocacy through actions that actors feel allow them to resist censorship of non-abusive, legal sexual expression.

\smallskip
\noindent
\textbf{Evaluative Criteria.}  
Society at large may wish to ensure that AIG-SC communities hold two criteria: 
(1) enablement of legal, free expressions of speech, and
(2) prevention of abuse. 
While our work provides insight into the tension between these two criteria, ways to concretely define and measure such criteria are beyond the scope of this work.

\smallskip
\noindent
\textbf{IAD Properties for Future Work.}  
While our work provides initial insight into how an IAD analysis may be defined and performed, much of the community-level properties of AIG-SC have yet to be concretely described.
These include, but are not limited to, understanding the action arena for collective, constitutional actions within AIG-SC %better defining how actions interact with one another, 
and examining exogenous variables including self-imposed community-based rules, implicit norms among the actors, and how these actions may change as AIG-SC technological abilities evolve.

%% file: appendices/methods-continued.tex
\section{Demographics (continued)}
\label{appendix:demographics}

\header{Age}
Almost half of participants (n=12) were within the 25-34 year old age group. Six participants were 18-24 and ten participants were above the age of 34. Table~\ref{table:age} displays the full breakdown.

\header{Occupational and Educational Background}
Participants were allowed to select multiple responses for their industry. The most common industries included gig work (9) science and technology (8) and arts, culture and entertainment (7). 
Exactly half of participants held a Bachelor's degree or higher. 
For the full statistics on both occupational background and educational attainment, see Tables~\ref{table:occupation} and ~\ref{table:education}.

\header{Race and ethnicity}
Twenty of the participants self-described as White,\footnote{Individuals who described themselves as ``White \& Hispanic'', ``Mexican caucasian'', ``presenting as white'' were included in both of these counts.} seven as Latin American,\footnote{Includes those who self-described as ``Latin American'', ``Hispanic'', ``Latino''.}, five as Asian, and two did not describe themselves. 

\header{Gender and sexuality}
Similarly, participants were given the option to self-describe their gender and sexual orientation. We categorized free-form responses to their closest representation. Our presentation of summary statistics therefore does not represent the full richness in participants' self-descriptions.

\header{Disability}
Participants were asked if they had a disability (including physical, mental health, neurodivergence). Fourteen participants responded "no", eleven participants responded "yes", and three preferred not to answer.

\input{appendices/demographics-tables}

%% file: appendices/demographics-tables.tex
\newpage

\small
\begin{table}[!h]
    \centering
  \begin{tabular}{l|c}
    \toprule
    How would you describe your sexual orientation? & \# Participants \\
    \midrule
    Heterosexual & 18  \\
    Bisexual/Gay/Lesbian/Pansexual/Queer & 9 \\
    No Response & 1 \\
    
    \toprule
    How would you describe your gender?\footnote{Several participants used gender and sex interchangeably e.g., reported their gender as `male', which we included in the count for men. We similarly do so for those who identify their gender as `female', who we include in the count for women.}  & \# Participants* \\
      \midrule
    Male/Man & 20 \\
    Non-Binary & 4 \\
    Female/Woman & 5 \\
    No Response & 1 \\
    \bottomrule
  \end{tabular}
    % \caption{Caption}
    \caption{Self-descriptions of sexual orientation and gender. * Several participants identify as multiple genders, and thus, these counts are not mutually exclusive.}
    \label{table:demographics}
\end{table}

\begin{table}[h!]
\centering
\begin{tabular}{|l|c|}
\hline
\textbf{Age Group} & \textbf{\# Participants} \\
\hline
18--21 years old & 2 \\
22--24 years old & 4 \\
25--34 years old & 12 \\
35--44 years old & 7 \\
45--54 years old & 2 \\
55+ years old & 1 \\
\hline
\end{tabular}
\caption{Number of participants per age group.}
\label{table:age}
\end{table}

\begin{table}[!ht]
\centering
\begin{tabular}{|l|c|}
\hline
\textbf{Category} & \textbf{Count} \\
\hline
Gig work & 9 \\
Science and technology & 8 \\
Arts, culture and entertainment & 7 \\
Other & 6 \\
Communications & 5 \\
Education & 4 \\
Business, management and administration & 3 \\
Sales & 3 \\
Health and medicine & 3 \\
Sex work (e.g., camming, stripping, escorting, etc.) & 2 \\
Architecture and engineering & 2 \\
Installation, repair and maintenance & 2 \\
\hline
\end{tabular}
\caption{Occupational background of participants.}
\label{table:occupation}
\end{table}

\begin{table}[h!]
\centering
\begin{tabular}{|l|c|}
\hline
\textbf{Education Level} & \textbf{Count} \\
\hline
Did not finish high school & 1 \\
High school graduate (diploma or equivalent) & 4 \\
Some college or Associate degree (e.g., A.A., A.S.) & 7 \\
Trade, technical, or vocational training & 2 \\
Bachelor's degree (e.g., B.A., B.S., B.Eng.) & 9 \\
Postgraduate degree (e.g., Master's, J.D., Ph.D.) & 5 \\
\hline
\end{tabular}
\caption{Educational background of participants.}
\label{table:education}
\end{table}